\documentclass[nofootinbib,reprint,superscriptaddress,preprintnumbers]{revtex4-2}

\makeatletter 
    
\renewcommand\onecolumngrid{
\do@columngrid{one}{\@ne}%
\def\set@footnotewidth{\onecolumngrid}
\def\footnoterule{\kern-6pt\hrule width 1.5in\kern6pt}%
}

\renewcommand\twocolumngrid{
        \def\footnoterule{
        \dimen@\skip\footins\divide\dimen@\thr@@
        \kern-\dimen@\hrule width.5in\kern\dimen@}
        \do@columngrid{mlt}{\tw@}
}%

\makeatother    

\usepackage[bottom]{footmisc}
\usepackage{amsmath,amssymb,marginnote}
\usepackage{multirow} 
\usepackage{booktabs} 
\usepackage{graphicx,subfigure}
\usepackage{dcolumn}
\usepackage{bm}
\usepackage[mathlines]{lineno}
\usepackage[colorlinks=true, allcolors=blue]{hyperref}
\usepackage[usenames]{color}
\usepackage{xcolor}
\newcommand{\liu}[1]{\textcolor{black}{#1}}

\begin{document}
\preprint{LA-UR-24-32057}
\title{Quasi-steady evolution of fast neutrino-flavor conversions}
\author{Jiabao Liu}
\affiliation{Department of Physics and Applied Physics, School of Advanced Science \& Engineering, Waseda University, Tokyo 169-8555, Japan}
\author{Hiroki Nagakura}
\affiliation{Division of Science, National Astronomical Observatory of Japan, 2-21-1 Osawa, Mitaka, Tokyo 181-8588, Japan}
\author{Masamichi Zaizen}
\affiliation{Department of Earth Science and Astronomy, The University of Tokyo, Tokyo 153-8902, Japan}
\author{Lucas Johns}
\affiliation{Theoretical Division, Los Alamos National Laboratory, Los Alamos, NM 87545 USA}
\author{Ryuichiro Akaho}
\affiliation{Department of Physics and Applied Physics, School of Advanced Science \& Engineering, Waseda University, Tokyo 169-8555, Japan}
\author{Shoichi Yamada}
\affiliation{Department of Physics, School of Advanced Science \& Engineering, Waseda University, Tokyo 169-8555, Japan}
\affiliation{Research Institute for Science and Engineering, Waseda University, Tokyo 169-8555, Japan}
\begin{abstract}
In astrophysical environments such as core-collapse supernovae (CCSNe) and binary neutron star mergers (BNSMs), neutrinos potentially experience substantial flavor mixing due to the refractive effects of neutrino self-interactions. Determining the survival probability of neutrinos in asymptotic states is paramount to incorporating flavor conversions' effects in the theoretical modeling of CCSN and BNSM. Some phenomenological schemes have shown good performance in approximating asymptotic states of fast neutrino-flavor conversions (FFCs), known as one of the collective neutrino oscillation modes induced by neutrino self-interactions. However, a recent study showed that they would yield qualitatively different asymptotic states of FFC if the neutrino number is forced to evolve. It is not yet fully understood why the canonical phenomenological models fail to predict asymptotic states. In this paper, we perform detailed investigations through numerical simulations and then provide an intuitive explanation with a quasi-homogeneous analysis. Based on the analysis, we propose a new phenomenological model, in which the quasi-steady evolution of FFCs is analytically determined. The model also allows us to express the convolution term of spatial wave number as a concise form, which corresponds to useful information on analyses for the non-linear feedback from small-scale flavor conversions to large-scale ones. Our model yields excellent agreement with numerical simulations, which lends support to our interpretation.
\end{abstract}
\maketitle
\section{Introduction}
In dense neutrino gases, neutrino self-interaction \cite{PANTALEONE1992128,PhysRevD.48.1462} can induce collective flavor conversions much faster than the neutrino oscillation in vacuum~\cite{PhysRevD.53.5382} and other relevant physical timescales in core-collapse supernovae (CCSNe) and binary neutron star mergers (BNSMs). The fast flavor conversion (FFC) represents such a case that has been attracting much attention in CCSN and BNSM modelers 
\liu{\cite{PhysRevD.110.103019,Shalgar_2024,xiong2024robustintegrationfastflavor,PhysRevD.109.123008,Froustey:2024ln,abbar2024applicationneuralnetworksreconstruction,nagakura2024neutronstarkickdriven,zaizen2024fastneutrinoflavorswaphighenergy,abbar2024detectingfastneutrinoflavor,nagakura2023basiccharacteristicsneutrinoflavor,nagakura2023globalfeaturesfastneutrinoflavor,PhysRevLett.131.061401,PhysRevD.107.103034,PhysRevLett.130.211401}}. Linear stability analysis of flavor conversions suggests that fast flavor instabilities (FFIs) commonly occur in CCSNe and BNSMs \cite{PhysRevD.100.043004,Nagakura_2019,PhysRevD.101.023018,PhysRevD.101.043016,PhysRevD.101.063001,Abbar_2020,PhysRevD.103.063013,PhysRevD.104.083025,Harada_2022,Akaho_2023,PhysRevD.99.103011,PhysRevD.101.023018,PhysRevD.101.043016,PhysRevD.100.043004,PhysRevD.104.083025,Harada_2022,Akaho_2023}, motivating detailed studies of the non-linear dynamics of FFCs \cite{PhysRevD.101.043009,PhysRevD.102.103017,PhysRevD.104.103003,PhysRevLett.128.121102,PhysRevD.104.103023,PhysRevD.106.103039,PhysRevLett.129.261101,PhysRevD.107.103022,PhysRevD.107.123021,xiong2023evaluating} including their interplay with various weak processes involving baryons and leptons \cite{PhysRevD.103.063001,PhysRevD.105.043005,PhysRevD.106.043031,PhysRevD.103.063002,Kato_2021,10.1093/ptep/ptac082,PhysRevD.105.123003,Kato_2022,PhysRevD.106.103031,PhysRevD.103.063002,PhysRevLett.122.091101}. 

Numerical simulations of quantum kinetic neutrino transport are the most straightforward way to study the non-linear phase of FFCs. However, there is a huge disparity between the scale of FFCs and the global system of CCSNe and BNSMs, which makes global simulations of FFCs intractable. For this reason, most previous CCSN and BNSM simulations incorporating effects of neutrino flavor conversions have been done by setting occurrence conditions and vigor of flavor conversions in parametric manners \cite{PhysRevD.107.103034,PhysRevLett.131.061401,PhysRevLett.126.251101,PhysRevD.105.083024,PhysRevD.106.103003}. This approach offers qualitative insights into how the dynamics of CCSN and BNSM are influenced by flavor conversions, but the prescription is not developed self-consistently with neutrino quantum kinetics, which would yield unrealistic outcomes.

Subgrid or coarse-grained models are representative frameworks to fill the gap between these phenomenological approaches and first-principle ones in solving the quantum kinetic equation (QKE). By taking advantage of the scale difference between neutrino flavor conversions and astrophysical systems, one of the authors in the present paper proposed a coarse-grained formulation, the so-called miscidynamics \cite{johns2023thermodynamicsoscillatingneutrinos}. In this method, QKE of neutrino transport is approximated based on the assumption that flavor conversions instantaneously settle down onto local equilibrium. Another (but relevant to miscidynamics; see \cite{johns2024subgridmodelingneutrinooscillations}) approach has also been proposed recently, which is Bhatnagar-Gross-Krook (BGK) subgrid model \cite{PhysRevD.109.083013}. The subgrid model essentially corresponds to a relaxation-time approximation. In this approach, we need to determine asymptotic states and timescales of flavor conversions one way or another, and then we replace the term of neutrino oscillation with a relaxation term towards asymptotic states of flavor conversions. A similar approach has also been proposed in \cite{xiong2024robustintegrationfastflavor}, and their global simulations showed an excellent capability to model the neutrino radiation field without solving QKE.

The accuracy of subgrid models depends to a large extent on how well one can determine asymptotic states of flavor conversions from coarse-grained neutrino distributions. There are numerous attempts to tackle the issue, in particular for FFCs; see, e.g., \cite{PhysRevLett.126.061302,PhysRevD.106.103039,PhysRevD.109.043024,richers2024asymptoticstatepredictionfastflavor,finalstate,PhysRevD.107.123021,xiong2023evaluating}. In \cite{finalstate,PhysRevD.107.123021}, we proposed that the stability and conserved quantities are key ingredients to determine the asymptotic states. More specifically, the asymptotic state of FFC needs to satisfy a condition that there is no crossing of ELN-XLN where ELN and XLN denote electron-neutrino and havey-leptonic-neutrino lepton number, respectively, in neutrino angular distributions. Given the condition of conserved quantities (which depends on the boundary condition of neutrino transport; see, e.g., \cite{PhysRevD.107.123021} for more details) and assuming fully depolarized in a certain angular region, we can approximately obtain the asymptotic state. Our method is consistent with both local- and global numerical simulations; see e.g., \cite{PhysRevD.104.103003,PhysRevD.106.043011} for the former and \cite{PhysRevLett.129.261101,PhysRevLett.130.211401} for the latter. We also note that our phenomenological model has been recently improved in \cite{xiong2023evaluating} and also extended to spatially multi-dimensional cases \cite{richers2024asymptoticstatepredictionfastflavor,george2024evolutionquasistationarystatecollective}.

Very recently, on the other hand, the authors in \cite{fiorillo2024fastflavorconversionsedge} found an intriguing phenomenon, in which FFCs can persist even after spatially-averaged ELN-XLN angular crossings disappear if the background neutrinos are forced to be evolved (see Sec.~\ref{sec:Motivation} for more details). This results in different asymptotic states of neutrinos from those predicted by the canonical asymptotic schemes mentioned above (since flavor conversions are assumed to be no longer active after the disappearance of ELN-XLN angular crossings). It should be mentioned that the authors in \cite{fiorillo2024fastflavorconversionsedge} attempted to give an interpretation of the \liu{observed FFC} with a quasi-linear theory, but it \liu{did not apply to the novel case}, exhibiting that the mechanism is still far from being understood. Since this anomaly of FFC evolution could be a significant obstacle in developing accurate subgrid models, this problem is worthy of in-depth investigation. In this paper, we scrutinize the specific (but important) problem. As we shall show later, this analysis offers new insights into the non-linear properties of self-induced flavor conversions.

This paper is organized as follows. After giving basic equations and notations in Sec.~\ref{sec:two-beam_basics}, we start with reviewing the result of \cite{fiorillo2024fastflavorconversionsedge} in Sec.~\ref{sec:Motivation}. In this section, we thoroughly explain why their results are interesting, which helps understand the backgrounds and motivations for readers unfamiliar with the literature. We then summarize our strategy to understand the phenomenon in Sec.~\ref{sec:Methods}. After describing essential information on our numerical schemes solving QKE, we summarize our numerical models which are designed to highlight key properties of quasi-steady evolution of FFC in Sec.~\ref{sec:Methods}. In Sec.~\ref{sec:Simu}, the numerical results are summarized and then we apply our quasi-homogeneous analysis to the results. Based on the analysis we construct an analytic model to determine the secular evolution of FFCs in Sec.~\ref{sec:semi_ana}. Finally, we conclude the paper by summarizing and discussing key findings in the present study.

Throughout the paper, we work in natural units ($c=\hbar=1$ where $c$ and $\hbar$ denote the speed of light and reduced Planck constant, respectively), which allows us to measure the time, length, and neutrino number density by the unit of neutrino self-interaction potential. We, hence, set the self-interaction potential as unity without loss of generality.

\section{Basic equation of FFC under two-beam model}
\label{sec:two-beam_basics}
The neutrino two-beam model is one of the common approaches to understand the core of physical processes in complex dynamics of flavor conversions. Following the previous study in \cite{fiorillo2024fastflavorconversionsedge}, we also work in the system with two neutrino beams streaming in opposite directions in a 1D box under a periodic boundary condition. One of the peculiar setups in the model compared to other studies of FFCs is that neutrinos are injected homogeneously into the simulation box. As we shall show later, the continuous neutrino injections affect spatially-averaged properties of FFCs even without ELN-XLN angular crossings (see Sec.~\ref{sec:Motivation} for more details). The QKE for such a system can be written with the polarization vector representation as
\begin{equation}
    \begin{split}
        (\partial_t+\partial_r)P_{R,0}&=C_R(t),\\
        (\partial_t-\partial_r)P_{L,0}&=C_L(t),\\(\partial_t+\partial_r)\mathbf{P}_R&=2\mathbf{P}_L\times\mathbf{P}_R+(0,0,C_R(t)),\\
        (\partial_t-\partial_r)\mathbf{P}_L&=-2\mathbf{P}_L\times\mathbf{P}_R+(0,0,C_L(t)).
    \end{split}
    \label{eq:QKE}
\end{equation}
In the expression, $t$ and $r$ denote time and spatial coordinates, respectively. $P_{L/R,\mu}$ represents the polarization vector for the left/right-going neutrino beam. It should be noted that we do not normalize the norm of $P_{L/R,\mu}$ as unity. The Greek letter $\mu$ denotes coordinate components in flavor space, which runs from 0 to 3. $C_{L/R}$ corresponds to the source term representing neutrino injection. We note that the injected neutrinos are assumed to be in flavor eigenstates, implying that the source term appears only in the zeroth and third component of QKE for $P_{L/R,\mu}$.

For developments of subgrid or coarse-grained models of flavor conversions, spatially averaged neutrino distributions are fundamental variables. The spatial average is defined for the arbitrary variable $f$ as
\begin{equation}
    \langle f(r)\rangle_V\equiv\int_0^L\frac{dr}{L}f(r),
    \label{eq:def_spave}
\end{equation}
where the angle brackets subscripted by V on the left side of the equation denote the spatial average. Hereafter, we use the capital letter $A$ for the spatial-averaged polarization vectors, i.e.,
\begin{equation}
    A_{L/R,\mu}\equiv\langle P_{L/R,\mu}\rangle_V.
    \label{eq:defA}
\end{equation}
From Eqs.~\ref{eq:QKE}~and~\ref{eq:defA}, the time evolution of $A_{L/R,\mu}$ can be expressed as,
\begin{equation}
    \begin{split}
        \partial_tA_{R,0}&=C_R(t),\\
        \partial_tA_{L,0}&=C_L(t),\\
        \partial_t\mathbf{A}_R&=2\langle \mathbf{P_L}\times\mathbf{P_R}\rangle_V+(0,0,C_R(t)),\\
        \partial_t\mathbf{A}_L&=-2\langle \mathbf{P_L}\times\mathbf{P_R}\rangle_V+(0,0,C_L(t)).
    \end{split}
    \label{eq:QKE_A}
\end{equation}
We note that the advection terms (which is involved on the left-hand side of Eq.~\ref{eq:QKE}) disappear due to periodic boundary conditions.

As we shall show in the following sections, there are two important quantities to characterize the quasi-steady evolution of the system. One of them is the cosine of the zenith angle of the polarization vectors in flavor space,
\begin{equation}
    \beta_{L/R}\equiv\frac{P_{L/R,3}}{|\mathbf{P}_{L/R}|}.
\label{eq:betadef}
\end{equation}
The other is the cosine of the angle between the left- and right-going polarization vectors
\begin{equation}
    \kappa\equiv\frac{\mathbf{P}_L\cdot\mathbf{P}_R}{|\mathbf{P}_L||\mathbf{P}_R|}.
    \label{eq:def_kappa}
\end{equation}
Before we delve into the details of our present work, we give an overview of the intriguing phenomenon found by \cite{fiorillo2024fastflavorconversionsedge} in the next section.

\section{Problem setup}
\label{sec:Motivation}
The authors in \cite{fiorillo2024fastflavorconversionsedge} studied the non-linear phase of FFCs in the two-beam model, in which they varied the neutrino injection (injection frequency and direction) among their numerical models. They demonstrated that the time evolution of FFCs depends on the neutrino injection, and they also showed that resultant asymptotic states are different among models. In some models, they found asymptotic states of FFCs to be consistent with our canonical picture, in which neutrino beams in either direction achieve complete depolarization and neutrino distributions propagating in the other direction can be determined by the conservation law \cite{finalstate}. However, there is a model (SR model in \cite{fiorillo2024fastflavorconversionsedge}) that shows the qualitatively different time evolution of FFCs. Understanding the anomaly observed in SR model is a subject of the present study.

In SR model, heavy-leptonic neutrinos ($\nu_x$) are initially set homogeneously in the neutrino beam with a left-going direction, and the norm of its polarization vector is set to be 0.5, i.e., $P_{L,1}=P_{L,2}=0,$ and $P_{L,3}=-0.5$. There are no neutrinos in the right-going direction at $t=0$, but electron-type neutrinos ($\nu_e$) are injected homogeneously with $C_R=1/240$ at $t \ge 0$ (see also Eq.~\ref{eq:QKE}), while no neutrinos are injected in the left-going direction until the end of the simulation (i.e., $C_L=0$). After adding a small perturbation at $t=0$, they investigated the time evolution of the system by solving QKE (Eq.~\ref{eq:QKE}). Since the $\nu_e$ injection in the right-going neutrinos generates ELN-XLN angular crossing, it triggers FFI. Once it enters into the non-linear phase, the spatially averaged $P_R$, i.e., $A_{R}$ (see Eq.~\ref{eq:defA}) nearly reaches a complete flavor depolarization ($\mathbf{A}_R \sim 0$). This is consistent with our canonical picture of FFC dynamics.

However, an intriguing phenomenon occurs after the time when the total amount of $\nu_e$ injection reaches the same as that of $\nu_x$ in the left-going beam at $t=0$. Hereafter, we refer to the critical time as $t_m$ ($t_m=0.5/C_R=120$ for SR model). One thing we do notice here is that $A_{L,3}$ becomes nearly zero at $t_m$. This can be understood from the conservation law of ELN and XLN \cite{finalstate,PhysRevD.104.103003}. It is important to note that in the phase of $t>t_m$, $\nu_e$ continues to be injected in the right-going direction, leading to $A_{R,3}>0$. In our canonical picture of FFC dynamics, flavor conversions are unable to change the global (or spatially-averaged) profiles of neutrino fields, since their ELN-XLN angular crossings have already disappeared. For this reason, the canonical scheme to determine asymptotic states predicts that the $\nu_e$ in the right-going neutrinos simply accumulates with time with sustaining $A_{L,3} \sim 0$. However, their simulations showed that the system evolves qualitatively differently from the prediction, in which both $A_{R,3}$ and $A_{L,3}$ increase with time, suggesting that FFCs keep shuffling flavors even after $t>t_m$. The asymptotic state of flavor conversions at $t=240$, which corresponds to the time when $\nu_e$ injection is stopped, is qualitatively different from other models.

\begin{figure}
    \centering
    \includegraphics[width=1.0\linewidth]{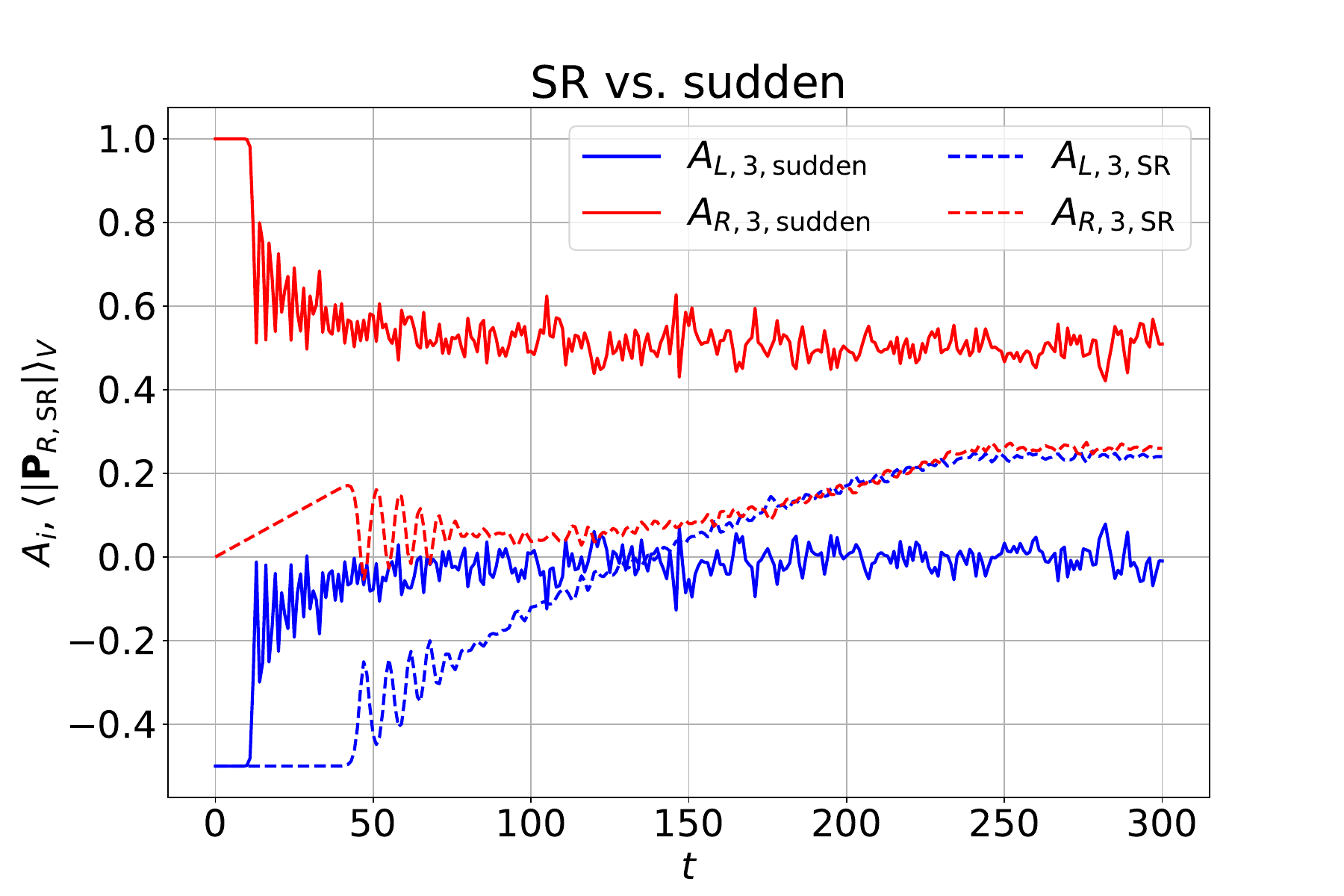}
    \caption{
    The time evolution of $A_{L,3}$ (blue) and $A_{R,3}$ (red) for SR (dashed lines) and sudden (solid lines) models.
    }
    \label{fig:SR_sudden}
\end{figure}

To see the overall trend of SR model more clearly, we display the numerical result in Fig.~\ref{fig:SR_sudden}. For comparison, we also show the result of the sudden model in \cite{fiorillo2024fastflavorconversionsedge}. It should be mentioned that these results are reproduced by our own QKE simulations; the details of our numerical simulations shall be described in Sec.~\ref{sec:Methods}. For the sudden model, the initial condition of left-going neutrinos is the same as that in SR model, whereas $\nu_e$ is also set at $t=0$ in the right-going neutrinos. The total amount of $\nu_e$ is twice of $\nu_x$ propagating in the left-going direction. As shown in the solid lines in Fig.~\ref{fig:SR_sudden}, neutrinos in both directions undergo strong flavor conversions at $t \sim 10$, while the flavor conversion for left-going neutrinos saturates with $A_{L,3} \sim 0$, indicating that the left-going neutrinos have undergone almost complete kinematic decoherence. For the right-going neutrinos, on the other hand, $A_{R,3}$ saturates at $\sim 0.5$, which is consistent with the canonical model of our asymptotic scheme \cite{finalstate}. As clearly seen in Fig.~\ref{fig:SR_sudden}, however, the asymptotic states of $A_{L/R,3}$ in SR model are very different from those in the sudden model. As described above, $A_{L,3}$ increases nearly constant in time until $\nu_e$ injection is ceased ($t=240$). We also find that both $A_{L,3}$ and $A_{R,3}$ co-evolve with time; in fact the two lines are nearly overlapped at $t>150$. 

One may expect that $A_{L,3} \sim A_{R,3}$ is a key property to solve the puzzle. However, we will show in the following section that this is just a coincidence. Instead, we provide more fundamental quantities to understand the physical process of the peculiar phenomenon. In the next section, we describe our strategy of how to understand these anomalies in FFCs.

\section{Methods and Models}
\label{sec:Methods}

\begin{table*}
    \centering
    \begin{tabular}{@{} p{2cm} p{8cm} @{}}
        \toprule
        \textbf{Model} & \textbf{Source Terms} \\ 
        \midrule
        SR & $C_R(t)=\frac{1}{240}$ and $C_L(t) = 0$ for $0 \leq t \leq 240$ \\
        \midrule
        SSR & $C_R(t)=\frac{1}{2400}$ and $C_L(t) = 0$ for $0 \leq t \leq 2400$ \\
        \midrule
        SSSR & $C_R(t)=\frac{1}{24000}$ and $C_L(t) = 0$ for $0 \leq t \leq 24000$ \\
        \midrule
        \multirow{2}{*}{SSSRs1} & $C_R(t)=\frac{1}{24000}$ and $C_L(t) = 0$ for $0 \leq t \leq 15000$, \\ 
         & $C_R(t)=0$ and $C_L(t) =\frac{1}{24000}$ for $15000 < t \leq 24000$ \\
        \midrule
        SSSRx3 & $C_R(t)=\frac{1}{24000}$ and $C_L(t) = 0$ for $0 \leq t \leq 72000$ \\
        \bottomrule
    \end{tabular}
    \caption{Setups for each Model Used in this Paper.}
    \label{tab:models}
\end{table*}

Similar to the previous work in \cite{fiorillo2024fastflavorconversionsedge}, we study non-linear phases of FFCs in the two-beam model by numerically solving QKE (Eq.~\ref{eq:QKE}) by adopting our simulation code in \cite{PhysRevD.107.123021,PhysRevD.109.083031}. The 1D-box is covered by a uniform grid with $N_z=10000$ and the spatial flux is reconstructed by implementing the seventh-ordered weighted essentially non-oscillatory (WENO) scheme\,\cite{LIU1994200,GEORGE2023108588}. The time integration is worked by the fourth-ordered Runge-Kutta scheme with a fixed time step size of $\Delta t = C_{\mathrm{CFL}}\Delta z$, where the Courant-Friedrichs-Lewy number is set to be $C_{\mathrm{CFL}}=0.4$. At the beginning of the simulation, random perturbations are added in the first component of $\mathbf{P}_{L}$, while we also correct the third component so that $|\mathbf{P}_{L}|$ is constant in space, i.e.,
\begin{equation}
\mathbf{P}_{L}(r)=(|\mathbf{P}_{L}|\epsilon(r),0,|\mathbf{P}_{L}|\sqrt{1-\epsilon(r)^2}).
\end{equation}
We set $|\epsilon(r)|<10^{-6}$ for all models.

We investigate five models in this study, whose control parameters are summarized in Table~\ref{tab:models}. We note that the initial condition is identical among all models, but other control parameters such as ($C_{R/L}$) and the duration of neutrino injection are varied systematically. As shown in Table~\ref{tab:models}, we study slower neutrino injection models than SR one. As shall be shown later, these numerical setups illuminate a universal property of quasi-steady evolutions of FFCs. In SSR and SSSR models, $C_R$ (the simulation time) is $10$- and $100$ times lower (longer) than those in SR one, respectively. This setup guarantees that the total amount of $\nu_e$ injection is the same among the three models.

We add two more models, SSSRs1 and SSSRx3, offering valuable information on the quasi-steady evolution of FFCs. For the former, it is identical to SSSR model up to the time of $t=15000$, but we swap the $\nu_e$ injection from right- to left-going neutrinos. As we shall show later, the asymptotic state of SSSRs1 model is different from SSSR one, indicating that the direction of neutrino injection has a substantial impact on FFC (see also Sec.~\ref{subsec:quahomoana} for the physical reason). The latter model is the same as SSSR model except for the duration of $\nu_e$ injection. The duration of $\nu_e$ injection is 3 times longer than SSSR one (but the same $C_R$), and therefore the total amount of $\nu_e$ also becomes higher than SSSR one. In SSSRx3 model, We pay special attention to the late-phase evolution, and the result is used to check our phenomenological model. The details will be discussed in Sec.~\ref{sec:semi_ana}.

\section{Numerical simulations}
\label{sec:Simu}

\subsection{Basic features}
\label{subsec:basicfeatures}

\begin{figure*}
    \centering
    \subfigure[]{\includegraphics[width=0.49\textwidth]{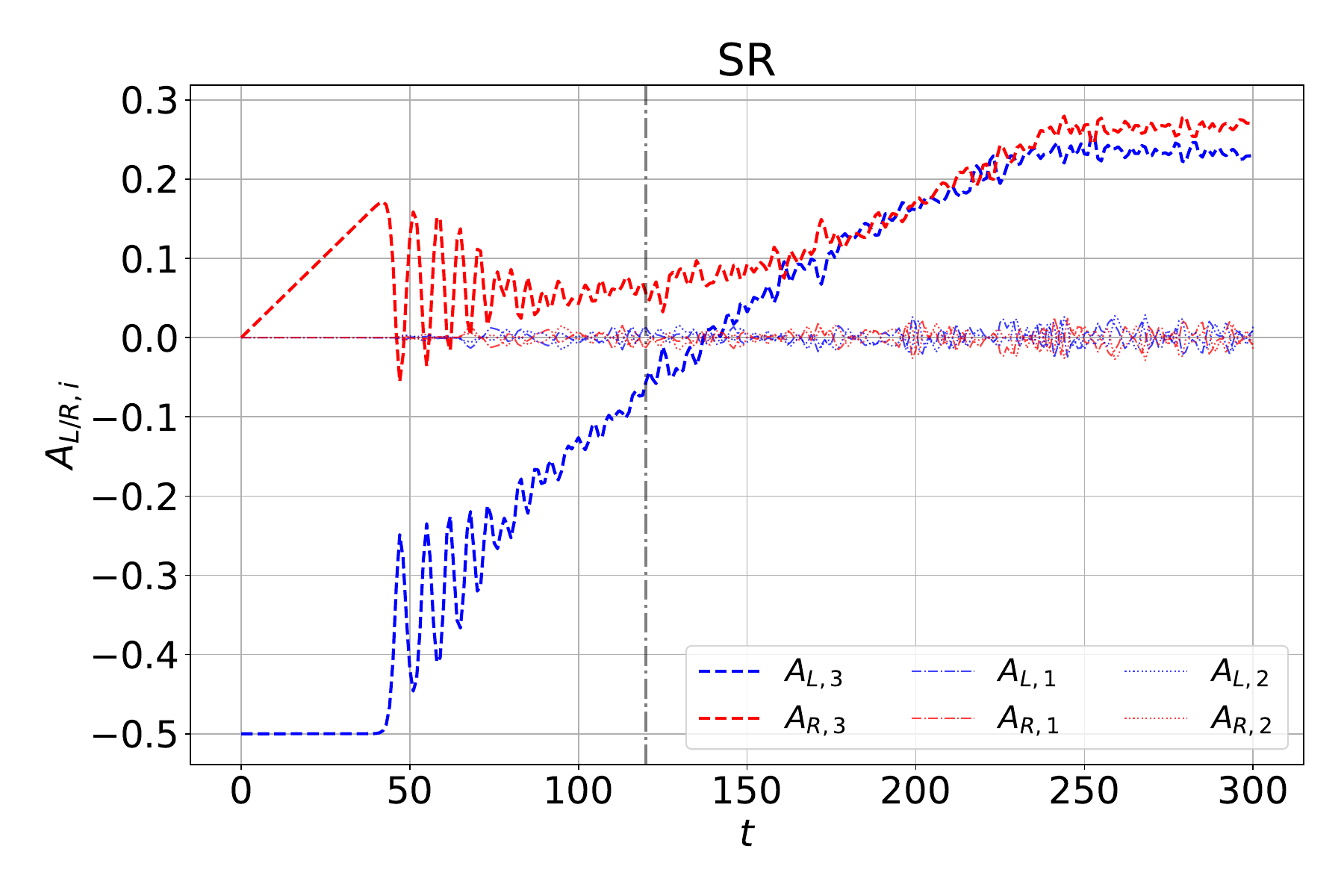}
         \label{subfig:SR_A}
     }
     \subfigure[]{\includegraphics[width=0.49\textwidth]{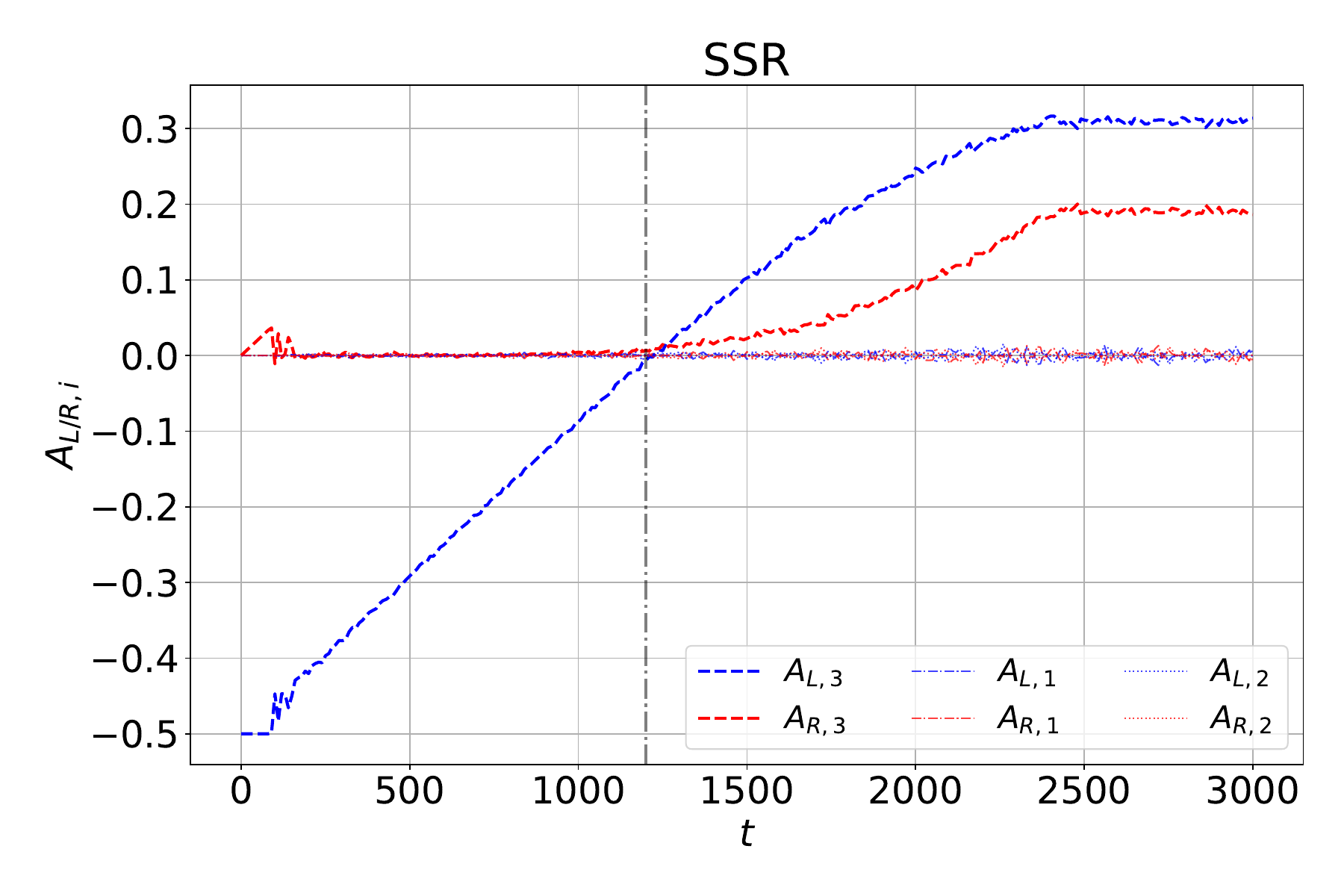}
         \label{subfig:SSR_A}
     }
     \subfigure[]{\includegraphics[width=0.49\textwidth]{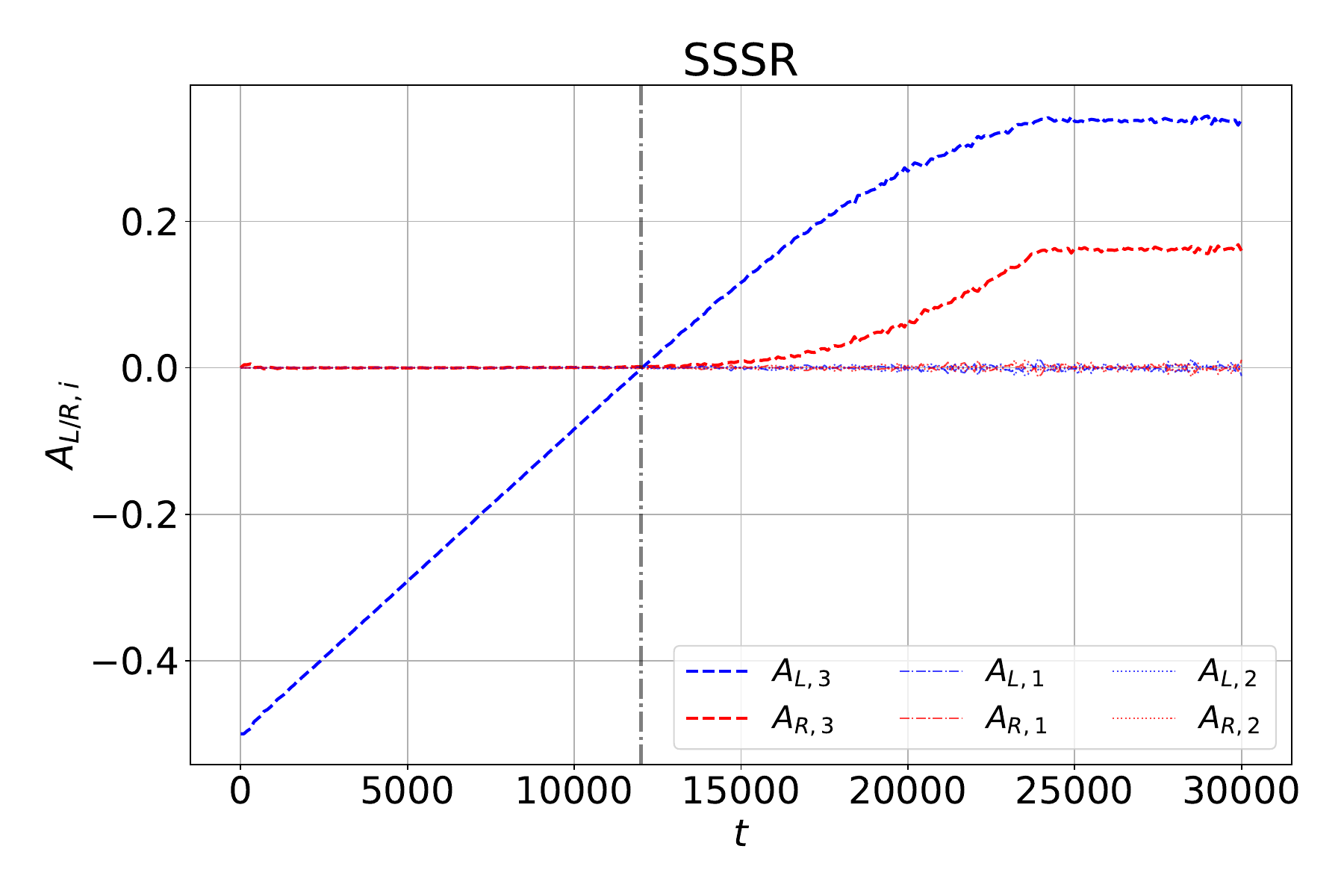}
         \label{subfig:SSSR_A}
     }
     \caption{The spatial-averaged polarization vector components $A_{i,\mu}$ for (a) SR model, (b) SSR model, and (c) SSSR models. For detailed model parameters, see Table.~\ref{tab:models}. The vertical dashed-dotted line marks $t=t_m$ (120 for SR, 1200 for SSR, 12000 for SSSR).
     }
     \label{fig:A}
\end{figure*}

\begin{figure}
    \centering
    \includegraphics[width=1.0\linewidth]{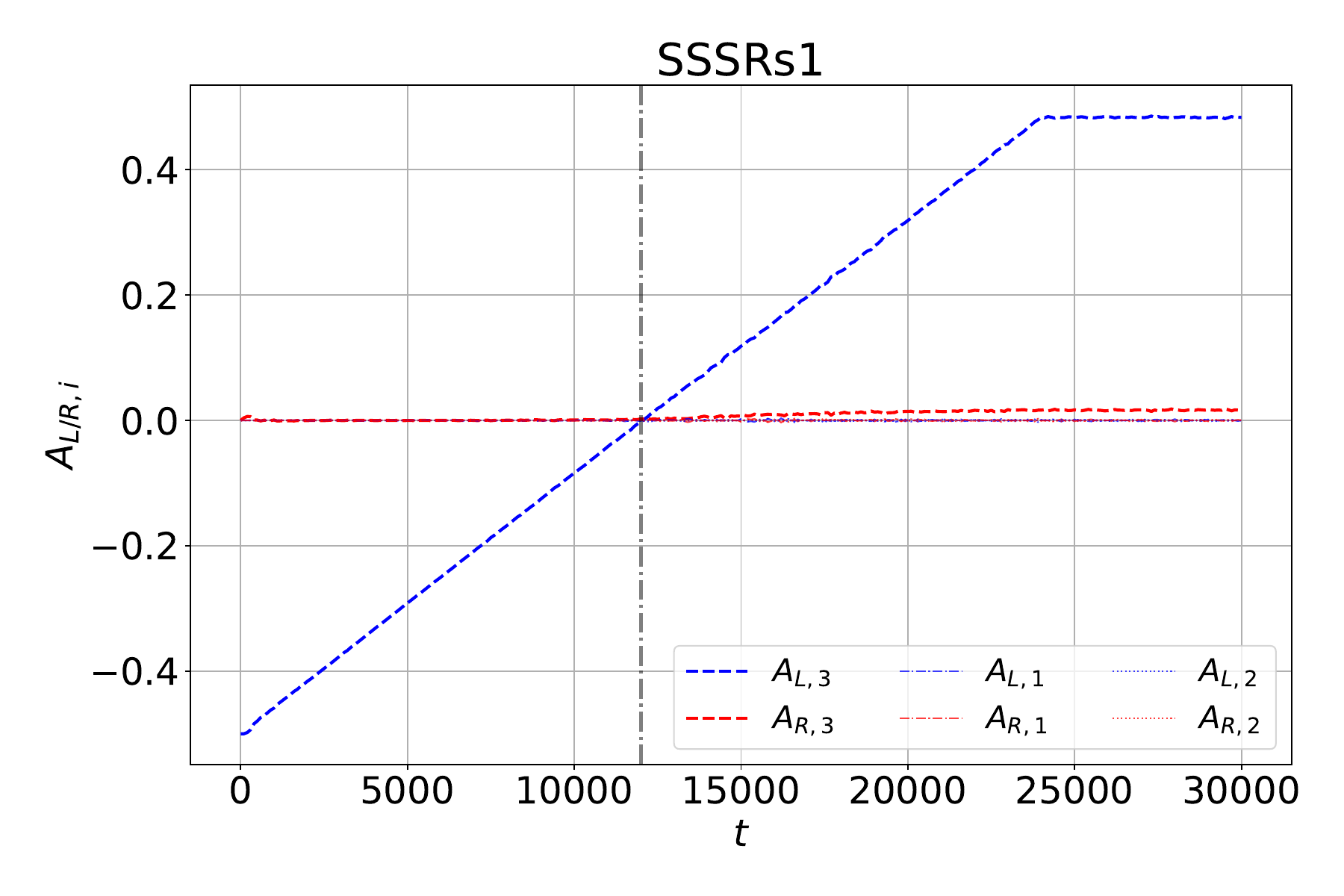}
    \caption{The same plots as in Fig.~\ref{fig:A} but for SSSRs1 model.}
    \label{fig:SSSRs1_basic}
\end{figure}

We first compare three models: SR, SSR, and SSSR. The time evolution of $A_{L/R,i}$, (where the Latin subscript, $i$, is used to denote the 1 to 3 components), and $\langle|\mathbf{P}_{L/R}|\rangle_V$ are shown in Fig.~\ref{fig:A}. At $t < t_m$, all models show a similar trend that $A_R$ tends to be depolarized and $A_L$ approaches zero. We also find that the temporal variability at the early phase becomes less vigorous for models with lower $C_R$ (or slower $\nu_e$ injection). This trend can be intuitively understood, since the disparity of time scales between FFC and $\nu_e$ injection becomes larger for lower $C_R$ models, exhibiting that the system evolves in more quasi-steady manners. It is also worthy of note that $A_{L,1}, A_{L,2}, A_{R,1}$, and $A_{R,2}$ are nearly zero in the entire time of simulations for all models.

Another noteworthy trend displayed in Fig.~\ref{fig:A} is that the time evolution of $A_{L/R,3}$ at $t>t_m$ in SSR model is very different from those in SR one, indicating that $A_{L,3} \sim A_{R,3}$ at $t>t_m$ in SR model is not a characteristic property in the quasi-steady evolution of FFC but just a coincidence. We also find that the difference between SSR and SSSR models is much smaller than the difference between SR and SSR, suggesting that the system converges to a certain asymptotic state by $C_R \to 0$ as long as the total amount of $\nu_e$ injection is the same. This exhibits that SSSR model is more appropriate as a reference model than SR one to investigate intrinsic properties of the quasi-steady evolution of FFC.

Figure~\ref{fig:SSSRs1_basic} shows the same quantities as Fig.~\ref{fig:A} but for SSSRs1 model. Let us recall that this model is the same as SSSR one up to $t=15000 (> t_m)$, but $\nu_e$ injection is swapped at that time from right- to left-going neutrinos. We are interested in whether or not the model follows the same trend as SSSR model. As shown in Fig.~\ref{fig:SSSRs1_basic}, $A_{R,3}$ remains nearly zero for the entire simulation, whereas $A_{L,3}$ vividly increases with time at $t>15000$. This result suggests that FFCs do not change the spatially averaged neutrino distributions. This exhibits that the dynamics is qualitatively different from SSSR model. We note that the result of SSSRs1 model is consistent with our canonical picture to determine the asymptotic state. In fact, the lack of ELN-XLN angular crossings in spatially-averaged neutrino radiation field does not give an impact on the asymptotic state of FFCs.

Below, we carry out more detailed analyses of these numerical results. As shown below, quasi-steady analysis can catch key characteristics of FFCs in quasi-steady evolution. The analysis offers key ingredients to develop a phenomenological model for the quasi-steady evolution of FFCs, which shall be discussed in Sec.~\ref{sec:semi_ana}.

\subsection{Quasi-homogeneous analysis}
\label{subsec:quahomoana}

\begin{figure}
    \centering
    \includegraphics[width=1.0\linewidth]{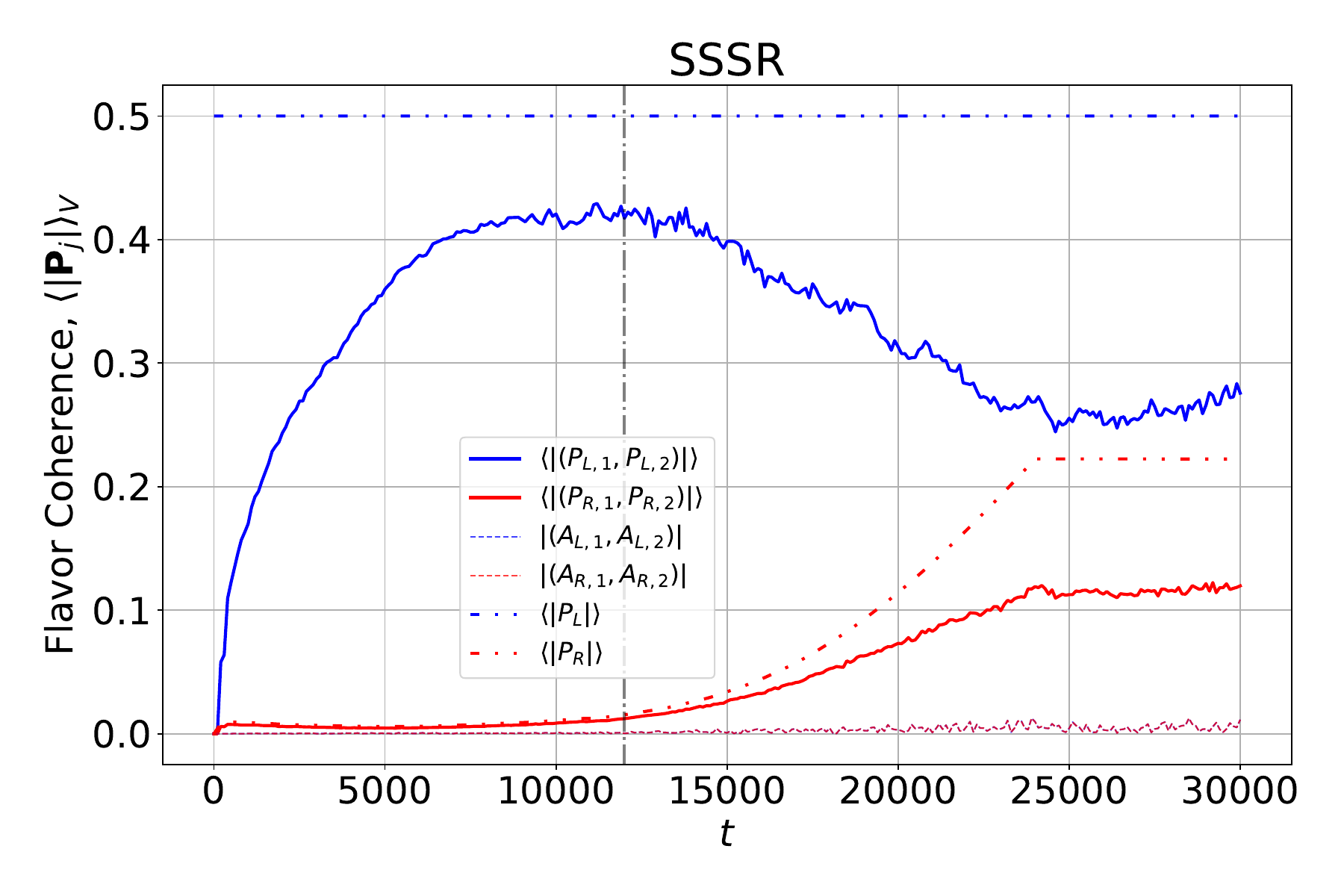}
    \caption{
    The time evolution of $\langle|\mathbf{P}_{L/R}|\rangle$, $\langle|\mathbf{P}^{T}_{L/R}|\rangle$, $\langle|\mathbf{A}^{T}_{L/R}|\rangle$, and $A_{R,3}$. The line type distinguishes these quantities, while the color distinguishes neutrino beams: blue (left-going neutrinos) and red (right-going neutrinos).
    }
    \label{fig:cohe}
\end{figure}

In discussing asymptotic states of FFCs with local QKE simulations, we have paid a special attention to $\mathbf{A}$, i.e., polarization vectors of homogeneous ($k=0$ where $k$ denotes the wave number) mode (see Eq.~\ref{eq:defA}). We note that such a homogeneous analysis has been working well. In fact, a canonical phenomenological model of asymptotic states of FFC was developed based on ELN-XLN angular distributions for $A_3$ (see, e.g., \cite{finalstate}). However, the analysis struggles to explain the quasi-steady evolution of FFCs summarized above, exhibiting the limitation of the canonical analysis. In this subsection, we propose another way to analyze our numerical results, which is hereafter referred to as the quasi-homogeneous analysis.

Before presenting the quasi-homogeneous analysis, let us highlight the weakness of canonical homogeneous analysis. It is, in general, impossible to judge only by homogeneous analysis whether the asymptotic states are in pure flavor eigenstates or have large flavor coherence. This problem can be seen in Figs.~\ref{fig:A}~and~\ref{fig:SSSRs1_basic}, in which the first and second components of $\mathbf{A}_{L/R}$ are nearly zero. As we shall show below, the degree of spatially averaged flavor coherency is a key ingredient to characterize the quasi-steady evolution of FFC, but such important information is smeared out in the homogeneous mode. The quasi-homogeneous analysis is designed to compensate for the weakness.

One of the representative variables in the quasi-homogeneous analysis is the norm of the polarization vector. Although the analysis with $\langle |\mathbf{P}_{L/R}| \rangle_V$ is not new (in fact, the authors in \cite{fiorillo2024fastflavorconversionsedge} also showed their spatial average in Fig.1), it is a good example to see the difference between the canonical- and quasi-homogeneous analyses. Let us consider the $|\mathbf{P}_{L}|$ in SSSR model. Since $|\mathbf{P}_{L}|$ is uniform in space at $t=0$ and no neutrinos are injected in the left-going direction ($C_L=0$), it does not evolve with time. This implies that the spatial average of $|\mathbf{P}_{L}|$, $\langle|\mathbf{P}_{L}|\rangle_V$ becomes constant in time; see also Fig.~\ref{fig:cohe}. It should be noted, on the other hand, that $|\mathbf{A}_L|$ is nearly equal to the norm of $A_{L,3}$ (since $A_{L,1} \sim A_{L,2} \sim 0$), indicating that $|\mathbf{A}_L|$ is a time-dependent quantity. The comparison between $|\mathbf{P}_{L}|$ and $|\mathbf{A}_L|$ clearly exhibits that they show different properties of polarization vectors (we note that they become identical if the system is homogeneous). This suggests that we can extract some hidden information on flavor coherence by comparing these quantities. See below for more details.

We start with the analysis with the norm of $\mathbf{P}_{L/R}$ and its associated quantities in the phase with $t<t_m$. The numerical results for SSSR model can be seen in Fig.~\ref{fig:cohe}. As already mentioned, $|\mathbf{P}_{L}|$ is constant in time, while we also find that the $|\mathbf{P}_{R}|$ remains to be nearly zero until $t=t_m$. The latter trend is obviously nontrivial since $\nu_e$ is injected in the right-going direction (i.e., $C_R$ is finite). A physical explanation is definitely needed. This trend can be analytically derived as follows. We first take an inner product of $\mathbf{P}_{R}$ with respect to the QKE for $\mathbf{P}_{R}$ in Eq.~\ref{eq:QKE}, which gives
\begin{equation}
    (\partial_t + \partial_z) \frac{|\mathbf{P}_R|^2}{2} = P_{R,3} C_R(t).
\label{eq:dt_PRnorm_orig}
\end{equation}
We, then, take the spatial average of Eq.~\ref{eq:dt_PRnorm_orig}, which results in dropping the advection term by using a periodic boundary condition, i.e.,
\begin{equation}
    \partial_t \frac{\langle|\mathbf{P}_R|^2\rangle}{2} = \langle P_{R,3} \rangle C_R(t).
\label{eq:dt_PRnorm_spave}
\end{equation}
As shown in Eq.~\ref{eq:dt_PRnorm_spave}, the increase rate of $\langle|\mathbf{P}_R|\rangle$ depends on not only $C_R$ but also $\langle P_{R,3} \rangle$. In SSSR model, $P_{R,3}$ (and $|\mathbf{P}_R|$) is zero at $t=0$ and then it increase with time with $\nu_e$ injection. In the very early phase (or linear phase of FFI), the norm grows linearly with time, $|\mathbf{P}_R|(t)=C_Rt$. Once the FFC enters the non-linear phase, $\mathbf{P}_R$ points in random directions in flavor space. This results in suppression of the increase of both $P_{R,3}$ and $|\mathbf{P}_R|$, which accounts for $|\mathbf{P}_{R}| \sim 0$ at $t<t_m$. Intuitively speaking, $\nu_e$ injection in the right-going neutrinos drives $\mathbf{P}_L$ dynamics so as to sustain $|\mathbf{P}_{R}| \sim 0$. As we shall show below, the flavor coherency for left-going neutrinos evolves with time, instead. On the other hand, $|\mathbf{P}_{R}|$ increases with time in the phase of $t>t_m$ until $\nu_e$ injection is ceased, which is because $P_{R,3}$ can no longer remain at zero; the detail will be discussed later.

In Fig.~\ref{fig:cohe}, the time evolution of the norm of the transverse component (or the first and second components) of $\mathbf{P}_{R/L}$ ($|\mathbf{P}^T_{L/R}|$) is also provided. The spatial average of $|\mathbf{P}^T_{L/R}|$ represents one of the key quantities that exhibits the overall degree of flavor coherency in the system. In fact, the degree of flavor coherency can be measured by $|\mathbf{P}^T_{L/R}|/|\mathbf{P}_{L/R}|$. As shown in the figure, $|\mathbf{P}^T_{L}|$ increases with time in the phase with $t<t_m$, and then it decreases afterward. This trend is consistent with our observation in $A_{L,3}$, which achieves the maximum flavor coherent state at $t=t_m$ ($\liu{\langle\beta_{L}\rangle_V} \sim 0$). It is worth noting that $|\mathbf{A}^T_{L/R}|$ is nearly zero during the entire simulation (which is also displayed in Fig.~\ref{fig:cohe}), exhibiting that \liu{the analysis based on only the homogeneous mode ($k=0$) is unable to quantify the overall degree of flavor coherency. The action of inhomogeneous modes ($k\neq0$) is important for the evolution of FFC.} 

\begin{figure}
    \centering
    \includegraphics[width=1.0\linewidth]{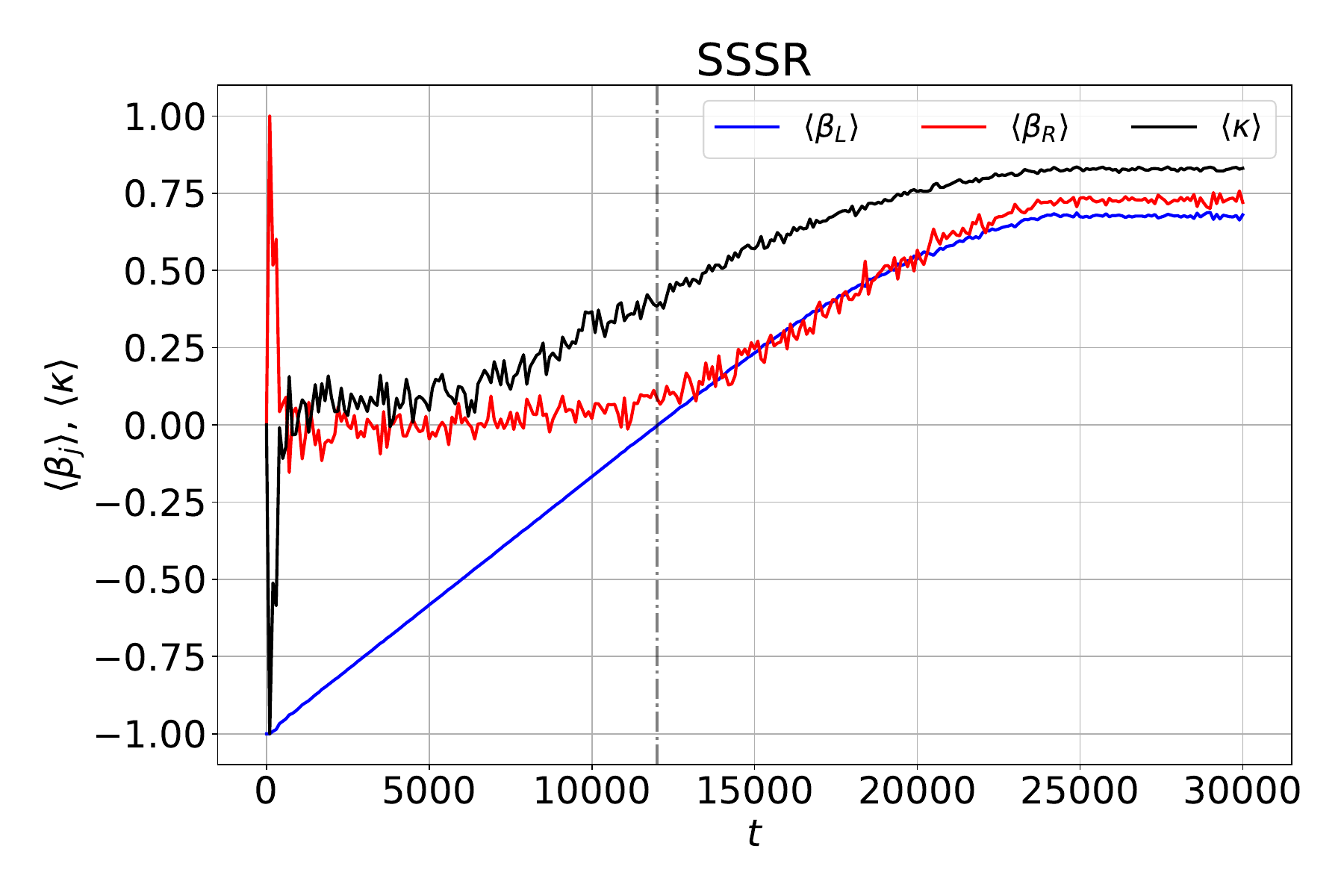}
    \caption{ The time evolution of
    $\langle\beta_{L}\rangle_V$ (blue line), $\langle\beta_{R}\rangle_V$ (red line), and $\langle\kappa\rangle_V$ (black line) for SSSR model.
    }
    \label{fig:beta}
\end{figure}

\begin{figure}
    \centering
    \includegraphics[width=1.0\linewidth]{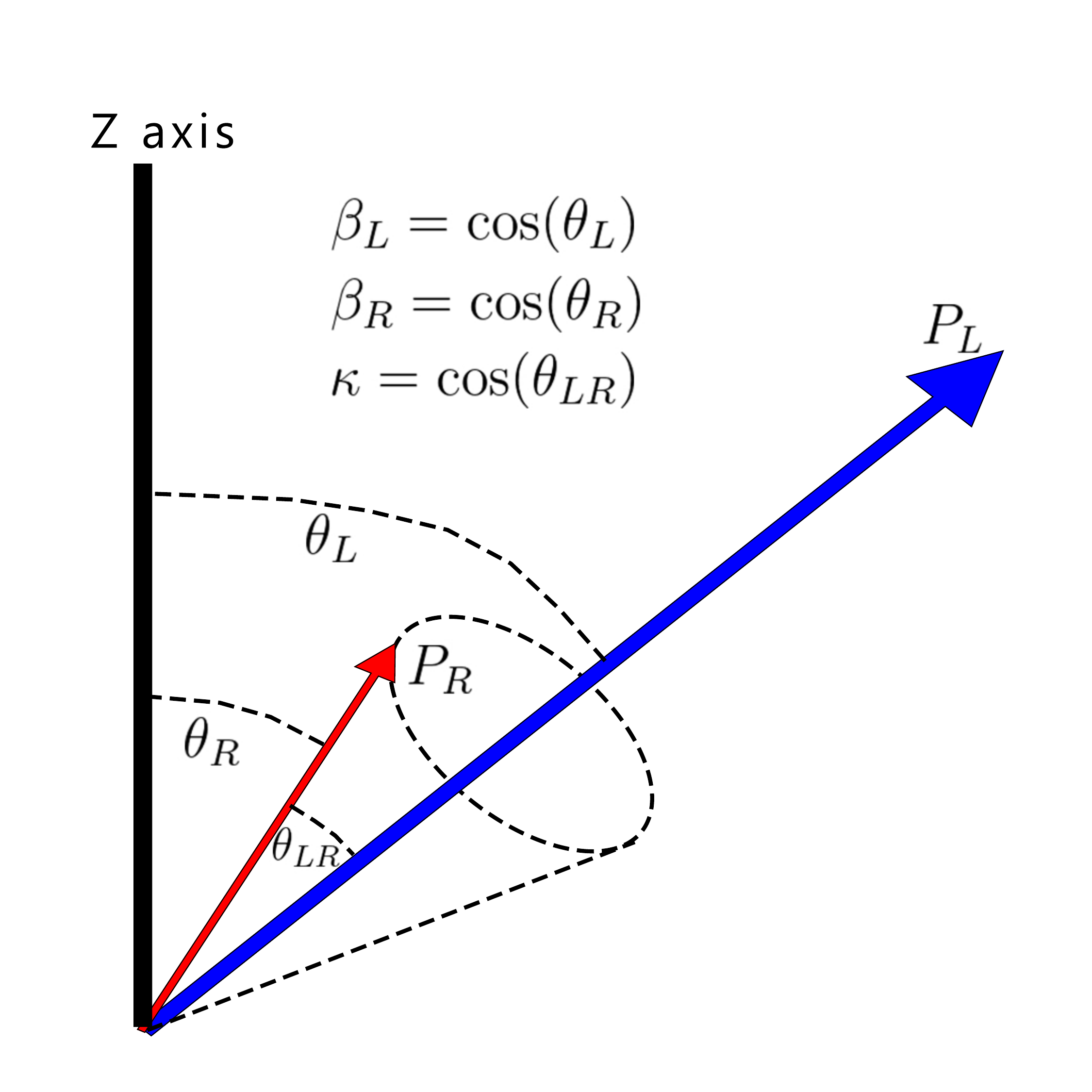}
    \caption{Schematic picture illustrating $\mathbf{P}_{L/R}$, $\beta_{L/R}$, and $\kappa$ under the quasi-homogeneous picture. \liu{The picture is relevant for $t>t_m$ where $\langle\kappa\rangle_V>0$ is established.}}
    \label{fig:Scheme}
\end{figure}

Below, we delve into the detail of the flavor coherency with $\beta$ and $\kappa$ (see Eqs.~\ref{eq:betadef}~and~\ref{eq:def_kappa}). It should be noted that $|\mathbf{P}^T_{L/R}|/|\mathbf{P}_{L/R}|$ can be rewritten in terms of $\beta_{L/R}$ (Eq.~\ref{eq:betadef}) as,
\begin{equation}
|\mathbf{P}^T_{L/R}|/|\mathbf{P}_{L/R}| = \sqrt{1-\beta_{L/R}^2},
\label{eq:flavcoh_beta}
\end{equation}
and the time evolution of $\langle|\mathbf{P}_R|\rangle_V$ can also be concisely expressed with $\langle\beta_R\rangle_V$,
\begin{equation}
    \partial_t\langle|\mathbf{P}_R|\rangle_V=\left\langle\frac{P_{R,3}}{|\mathbf{P}_R|}\right\rangle_VC_R(t)=\langle\beta_R\rangle_VC_R(t).
\label{eq:dt_PRnorm_1}
\end{equation}
We also note that $\beta_{L/R}$ is expected to provide some insights into $A_{L/R,3}$, since $A_{L/R,3}$ can be expressed as $\langle |\mathbf{P}_{L/R}| \beta_{L/R}\rangle_V$. Interestingly, our numerical results show that $|\mathbf{P}_{L/R}|$ and $\beta_{L/R}$ can be factorized, i.e.,
\begin{equation}
A_{L/R,3} 
=\langle|\mathbf{P}_{L/R}|\rangle_V\langle\beta_{L/R}\rangle_V.
\label{eq:factorize_beta}
\end{equation}
The success of factorization indicates that the dynamics of $|\mathbf{P}_{L/R}|$ and the degree of flavor coherency (represented by $\beta$) are determined independently. As we shall show below, $\langle \beta_{L/R} \rangle$ is a key variable to characterize the dynamics of the system.

As displayed in Fig.~\ref{fig:beta}, $\langle \beta_R \rangle_V$ and $\langle \beta_L \rangle_V$ have remarkably different time evolution in the phase at $t<t_m$. The former remains nearly zero, whereas the latter linearly increases with time. These dynamical properties can be understood through the canonical picture of the non-linear phase of FFC. As already mentioned, right-going neutrinos experience strong flavor conversions to erase ELN-XLN angular crossings. As a result, $\mathbf{P}_{R}$ points in random directions, which makes $\langle\beta_R\rangle_V \sim 0$. For left-going neutrinos, on the other hand, the flavor coherency increases with time by FFC while preserving $|\mathbf{P}_{L}|$. In our quasi-homogeneous picture, the polarization vector rotates in flavor space (see also Fig.~\ref{fig:Scheme} as a schematic picture to see relations among key variables\liu{, which is relevant for $t>t_m$ where $\langle\kappa\rangle_V>0$ is established}\footnote{\liu{Notably, $\langle\kappa\rangle_V$ begins to peel away from 0 long before $t_m$. Note that for $t<t_m$, although $|\mathbf{P}_R|$ is small, it is finite. Therefore, the value of $\langle\kappa\rangle_V$ in the early phase is meaningful. Although the dynamics is not yet fully understood, it is related to the local synchronization of the azimuthal angle between $\mathbf{P_L}$ and $\mathbf{P}_R$ in the nonlinear regime. In fact, there can exist an attractor near the state where the average azimuthal angle agrees between the two beams. The detailed analysis is postponed to our future work.}}), and the time evolution of directional cosine with respect to the flavor axis is represented by $\langle \beta_L \rangle_V$.

\begin{figure}
    \centering
    \includegraphics[width=1.0\linewidth]{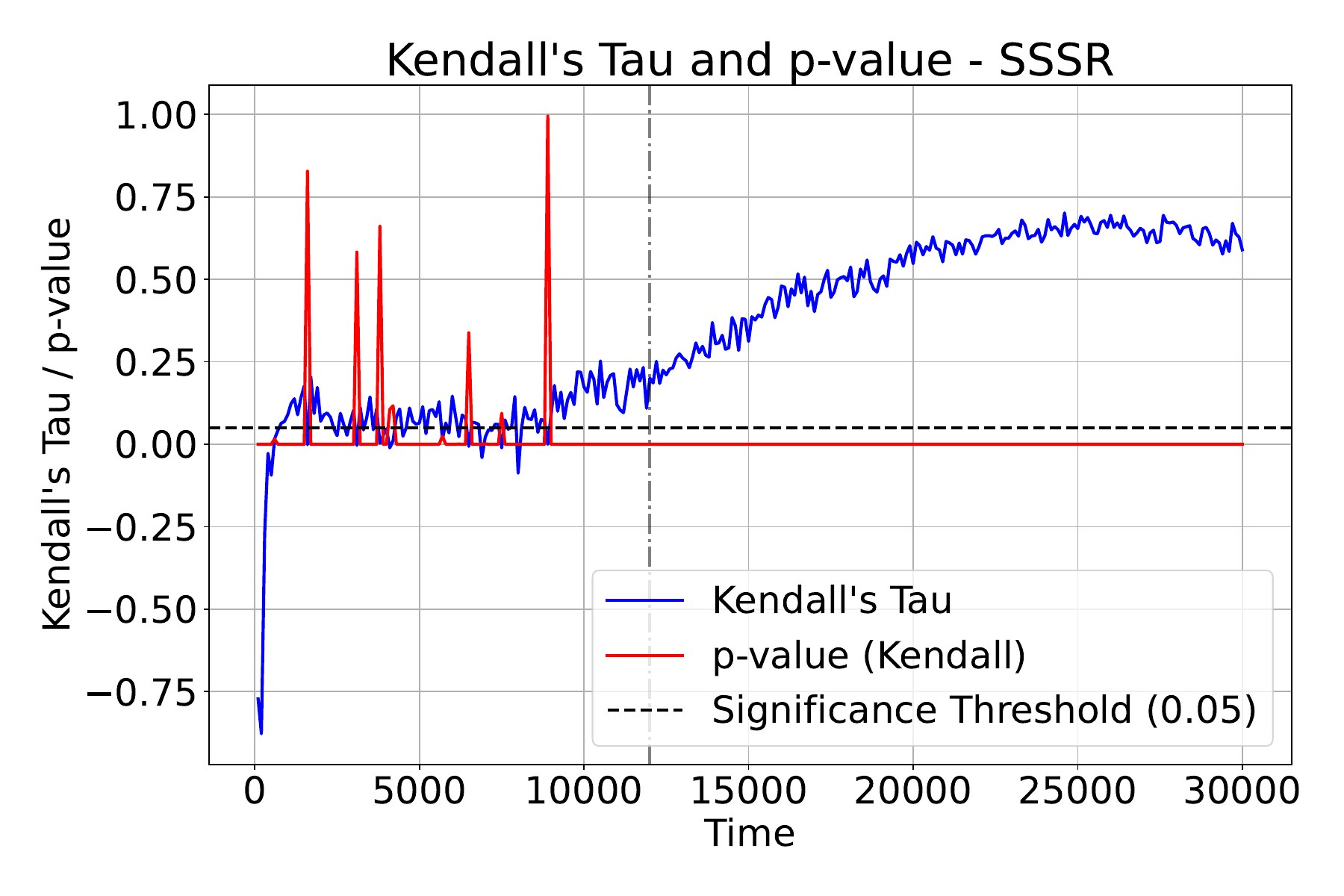}
    \caption{The correlation between the distributions of $\beta_L$ and $\beta_R$ quantified by the Kendall rank correlation coefficient. We also show the p-value and significance threshold as references.
    }
    \label{fig:SSSR_Corr}
\end{figure}

One of the most striking findings in this study is that $\langle \beta_{L} \rangle$ and $\langle \beta_{R} \rangle$ are nearly equal to each other in the phase of $t>t_m$ for SSSR model (see Fig.~\ref{fig:beta}), suggesting that the degrees of neutrino flavor polarization \liu{tend to get correlated on average} between the \liu{left-going beam and the right-going beam}. To see the correlation quantitatively, we compute the Kendall rank correlation coefficient concerning their spatial distributions. The time evolution of the coefficient is portrayed in Fig.~\ref{fig:SSSR_Corr}. As shown in the figure, they are essentially no correlation until $t=t_m$ (which would be due to the nearly complete depolarization of $\mathbf{P}_R$), but they are in strong positive correlation at $t>t_m$. This indicates that the repolarization of $P_R$ is dictated by that of the left-going neutrinos. 

\liu{Note that the correlation coefficient is not 1 even if $\langle\beta_L\rangle_V$ is nearly equal to $\langle\beta_R\rangle_V$. This can be understood from the schematic picture depicted in Fig.~\ref{fig:Scheme}. $P_R$ rotates around $P_L$, indicating that $\beta_L$ and $\beta_R$ are in general different locally. On the other hand, as shown in Fig.~\ref{fig:SSSR_Corr}, the correlation coefficient increase with time because $\kappa$ approaches 1. Furthermore, an increasing correlation coefficient confirms that $\beta_L$ and $\beta_R$ tend to be concordant at each spatial position, in addition to the co-evolution of $\langle\beta_L\rangle_V$ and $\langle\beta_R\rangle_V$.} 

As we shall show below, the dynamical property of $\kappa$ could account for the qualitative trend of $\langle \beta_L \rangle_V \sim \langle \beta_R \rangle_V$ at $t > t_m$. To see this, it is illustrative to derive the time evolution equation for $\langle \kappa \rangle_V$. Ignoring the advection term, we can obtain the following relation from Eqs.~\ref{eq:QKE}~and~\ref{eq:def_kappa},
\begin{equation}
  \begin{split}
    &\partial_t(\mathbf{P}_L\cdot\mathbf{P}_R)
    =\partial_t(|\mathbf{P}_L||\mathbf{P}_R|\kappa) \\
    &\sim P_{L,3}C_R(t) +P_{R,3}C_L(t),
  \end{split}
  \label{eq:forkappaevo_1}
\end{equation}
and then we take the spatial average as
\begin{equation}
    \partial_t(\langle|\mathbf{P}_L||\mathbf{P}_R|\kappa\rangle_V) \sim A_{L,3}C_R(t)+A_{R,3}C_L(t).
    \label{eq:forkappaevo_2}
\end{equation}
Assuming that spatial average of the left hand side of Eq.~\ref{eq:forkappaevo_2} can be factorized, i.e., 
\begin{equation}
    \langle|\mathbf{P}_L||\mathbf{P}_R|\kappa\rangle_V \sim \langle|\mathbf{P}_L|\rangle_V\langle|\mathbf{P}_L|\rangle_V\langle\kappa\rangle_V,
    \label{eq:kappafactorize}
\end{equation}
we can obtain the time evolution equation for $\langle \kappa \rangle_V$ as
\begin{equation}
\partial_t \langle \kappa \rangle_V \sim
\begin{cases}
    \frac{C_R(t)\left[\langle \beta_L \rangle_V - \langle \kappa \rangle_V \langle \beta_R \rangle_V \right]}{\langle |\mathbf{P}_R| \rangle_V}
    & \text{if } C_R \neq 0, C_L = 0 \\[0.5em]  
    \frac{C_L(t)\left[\langle \beta_R \rangle_V - \langle \kappa \rangle_V \langle \beta_L \rangle_V \right]}{\langle |\mathbf{P}_L| \rangle_V}
    & \text{if } C_R = 0, C_L \neq 0,
\end{cases}
\label{eq:tevokappa}
\end{equation}
while SSSR model corresponds to the upper case ($C_R \neq 0, C_L = 0$) in Eq.~\ref{eq:tevokappa}.

\begin{figure}
    \includegraphics[width=0.49\textwidth]{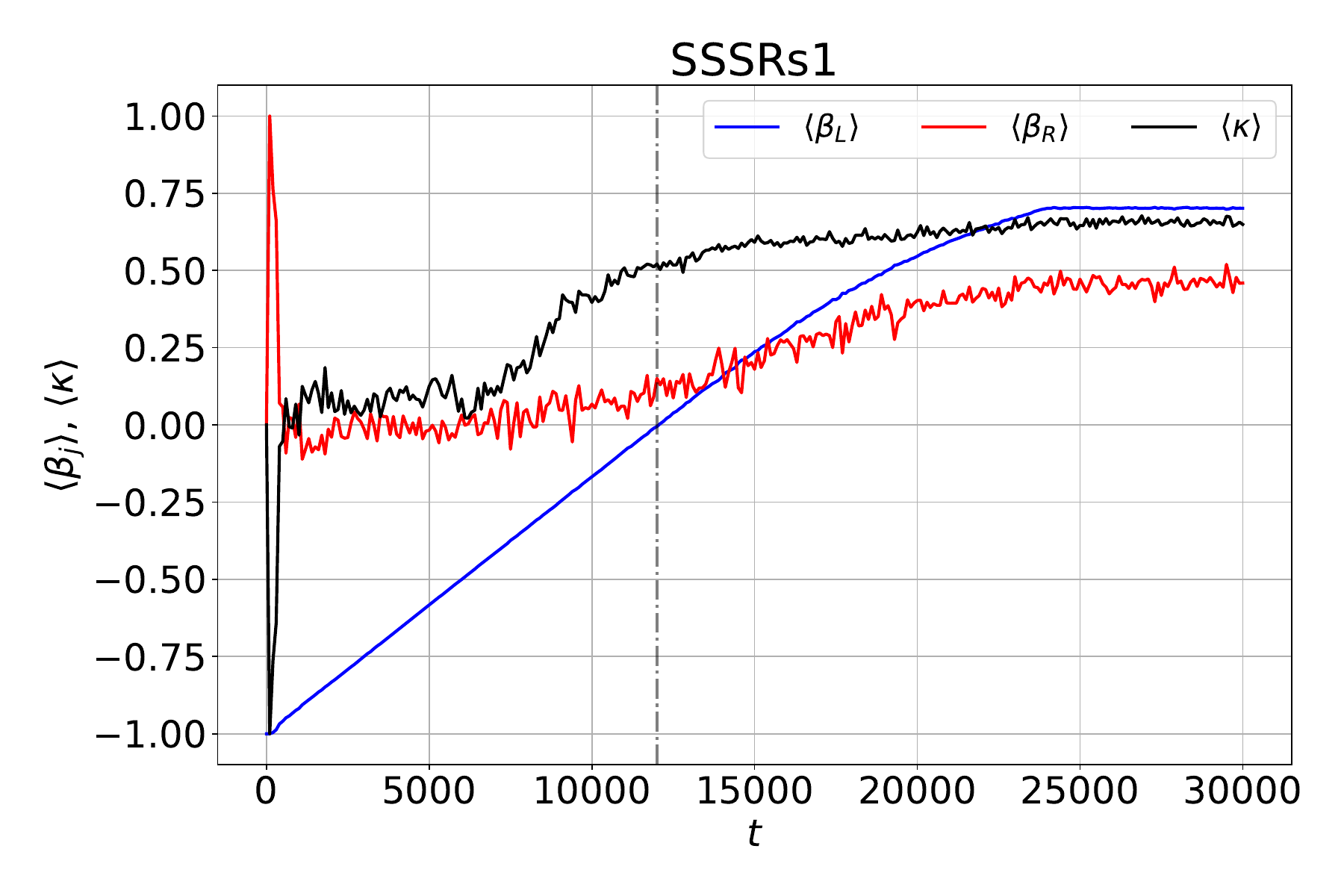}
     \caption{
     The same plots as in Fig.~\ref{fig:beta} but for SSSRs1 model.
     }
     \label{fig:SSSRs1_beta}
\end{figure}

Under the assumption that effects of neutrino advection are ignored, the system can be understood similar to a canonical homogeneous picture, in which both $\mathbf{P}_R$ and $\mathbf{P}_L$ rotate around the polarization vector for the zeroth-angular moment. Assuming $|\mathbf{P}_L| \gg |\mathbf{P}_R|$, the zeroth-angular moment is dominated by the left-going neutrinos, indicating that $\mathbf{P}_R$ rotates around $\mathbf{P}_L$ and $\kappa$ represents the opening angle (see Fig.~\ref{fig:Scheme}). At the time when both $\langle \beta_L \rangle_V$ and $\langle \beta_R \rangle_V$ becomes zero (which is around the time of $t_m$), $\langle \kappa \rangle_V$ is roughly $0.5$. At $t> t_m$, the term of $\langle \beta_L \rangle_V - \langle \kappa \rangle_V \langle \beta_R \rangle_V$ should be positive, indicating that $\langle \kappa \rangle_V$ increases with time. This exhibits that $\nu_e$ injection in the right-going direction brings $\mathbf{P}_R$ and $\mathbf{P}_L$ closer together (see also Fig.~\ref{fig:Scheme}), which accounts for $\langle \beta_L \rangle_V \sim \langle \beta_R \rangle_V$.

Let us make a few remarks here. First, one might think that $\langle \beta_L \rangle_V \sim \langle \beta_R \rangle_V$ can be obtained regardless of $\langle \kappa \rangle_V$ by our homogeneous picture (Fig.~\ref{fig:Scheme}), since $\mathbf{P}_R$ is assumed to rotate around $\mathbf{P}_L$. In fact, this statement is true in the case that both $\langle \beta_L \rangle_V$ and $\langle \beta_R \rangle_V$ are are nearly zero (which corresponds to the phase at $t \sim t_m$). However, $\langle \beta_R \rangle_V$ becomes, in general, different from $\langle \beta_L \rangle_V$ in cases with $\langle \beta_L \rangle_V > 0$. This can be understood in a limiting case with $\langle \beta_L \rangle_V \sim 1$. In such a case, we obtain $\langle \beta_R \rangle_V \sim \langle \kappa \rangle_V $, exhibiting that $\langle \kappa \rangle_V$ needs to be nearly unity for $\mathbf{P}_R$ and $\mathbf{P}_L$ to remain almost parallel. Next, Eq.~\ref{eq:tevokappa} also illustrates the physical picture of why SSSRs1 model (the case that $\nu_e$ injection is swapped at $t=15000$) has a qualitatively different trend from SSSR one. For SSSRs1 model, we can use the bottom equation in Eq.~\ref{eq:tevokappa}. The most important difference between the two cases is the denominator on the right-hand side of these equations. As mentioned already, $|\mathbf{P}_L| \gg |\mathbf{P}_R|$ is satisfied in both models, indicating that the dynamical time scale for $\langle \kappa \rangle_V$ in SSSRs1 model is much longer than SSSR one. In addition to this, $|\mathbf{P}_L|$ increases with time due to $\nu_e$ injection, which makes $\langle \kappa \rangle_V$ freeze out. We confirm the trend in our numerical simulations; the time evolution of $\langle \beta_{L/R} \rangle_V$ and $\langle \kappa \rangle_V$ is displayed in Fig.~\ref{fig:SSSRs1_beta}. As shown in the figure clearly, $\langle \kappa \rangle_V$ becomes nearly constant at $t > 15000$, and consequently $\langle \beta_{L/R} \rangle_V$ deviate from each other. This argument supports our claim that $\langle \kappa \rangle_V$ is another key factor in characterizing a quasi-steady evolution of FFC.

It should be mentioned that the above argument with $\langle \kappa \rangle_V$ is responsible only for the qualitative trend, and advection terms should play non-negligible roles for more quantitative arguments. In fact, the pure homogeneous system leads to qualitatively different results from the models considered in this study. Addressing this issue requires more systematic and detailed studies, which is therefore better to defer to another paper.

\section{Semi-Analytic Model}
\label{sec:semi_ana}

\begin{figure*}
    \centering
    \subfigure[]{\includegraphics[width=0.49\textwidth]{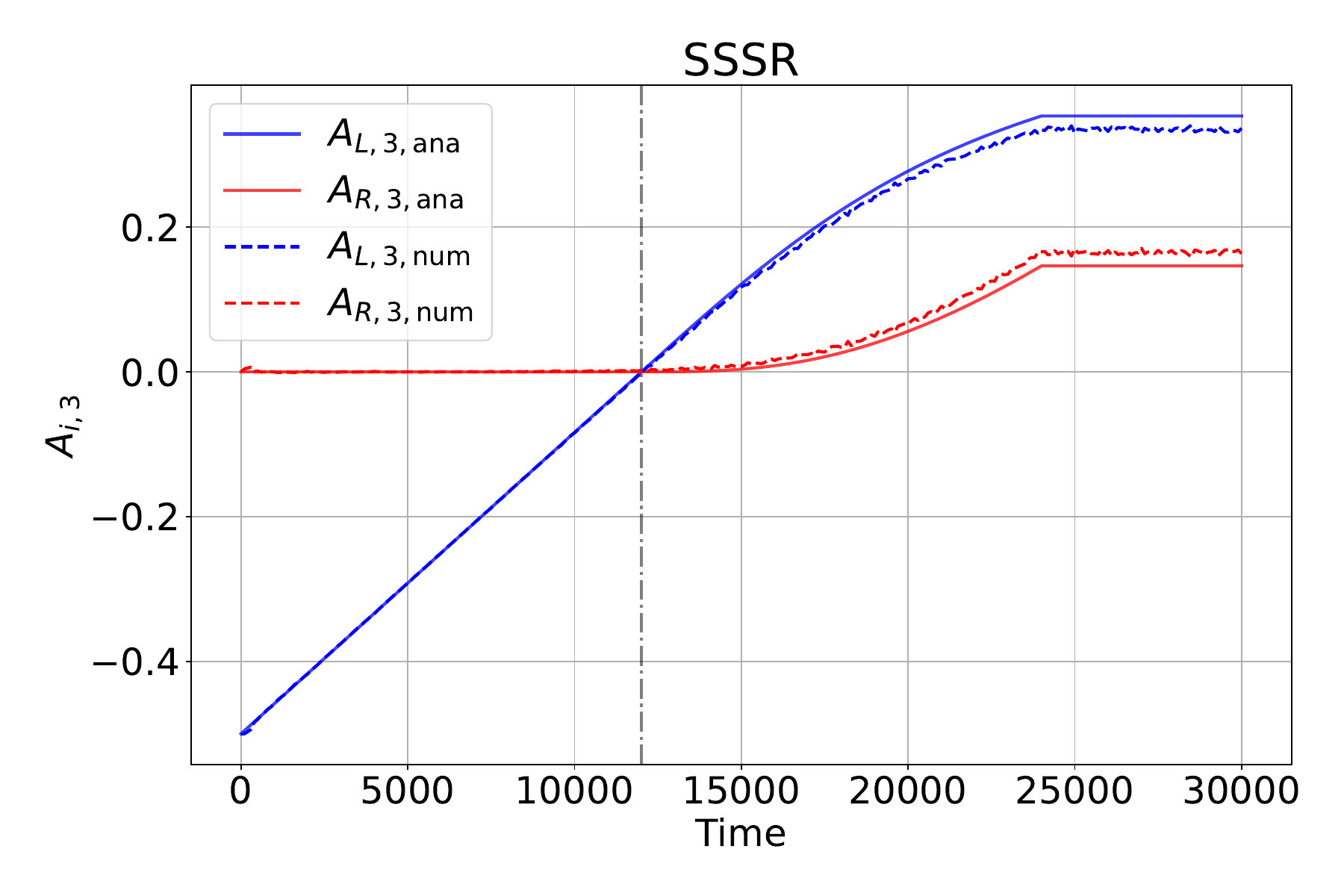}
         \label{subfig:semi_A}
     }
     \subfigure[]{\includegraphics[width=0.49\textwidth]{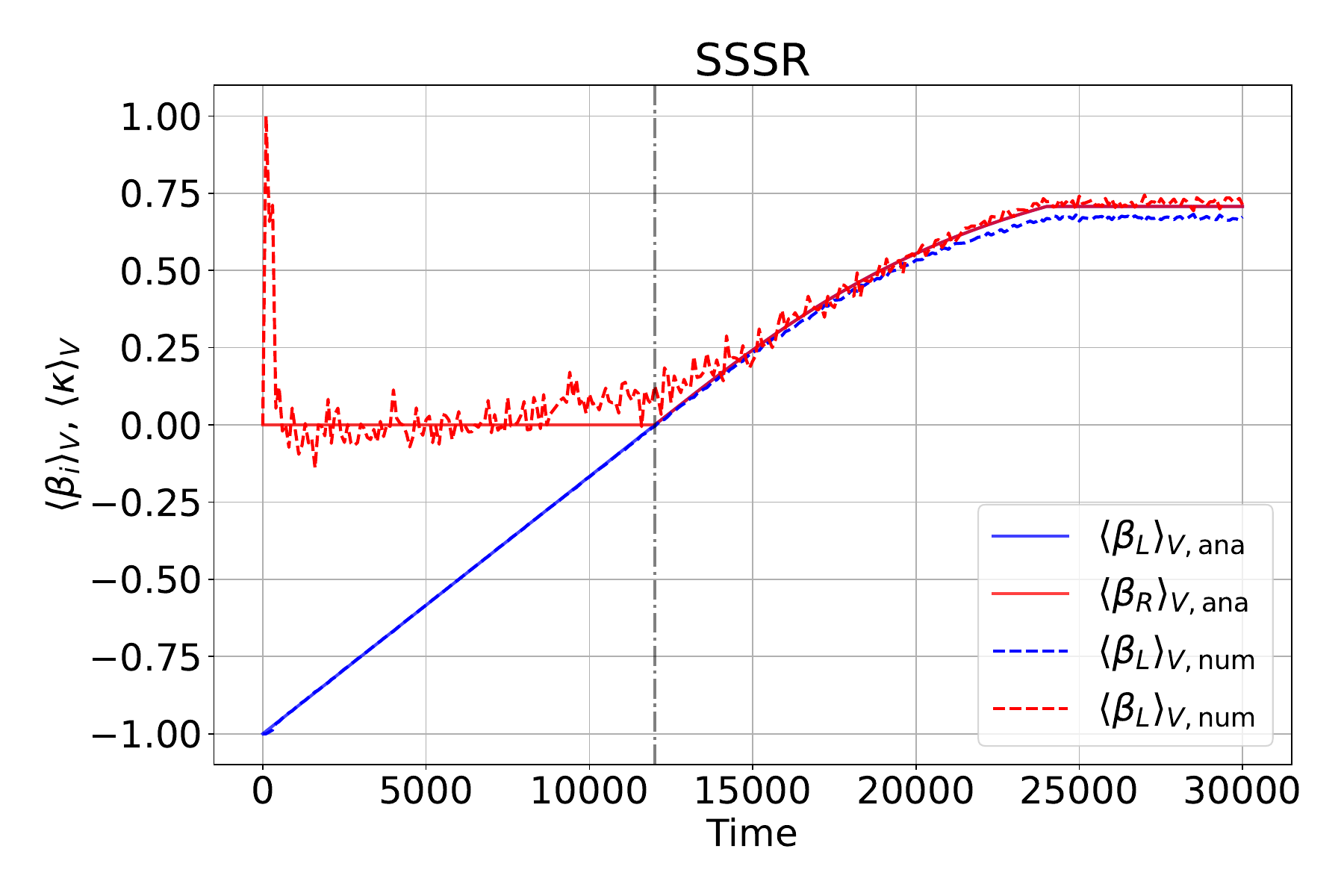}
         \label{subfig:semi_beta}
     }
     \subfigure[]{\includegraphics[width=0.49\textwidth]{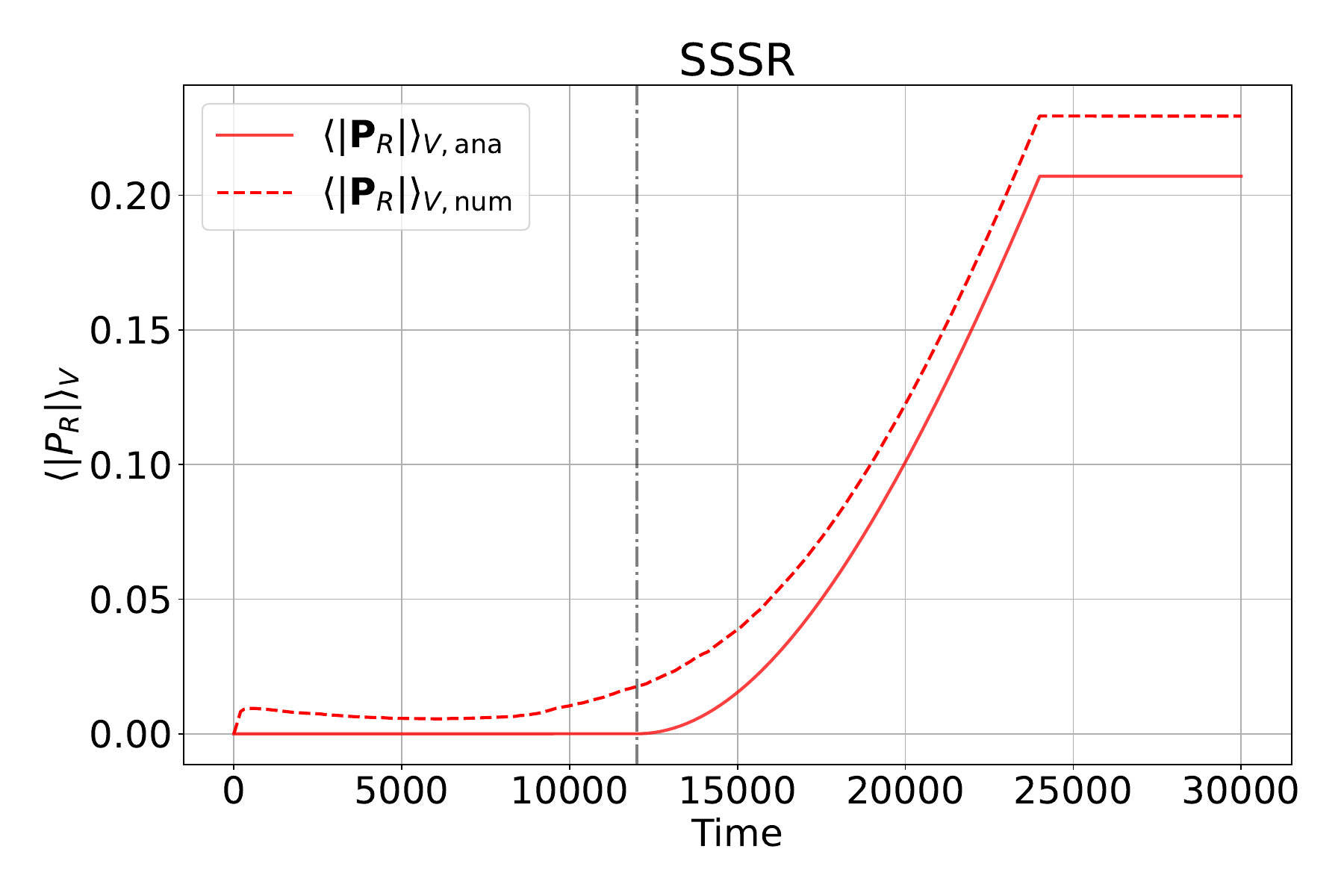}
         \label{subfig:semi_PR}
     }
     \caption{
     Comparing the time evolution of key quantities between our phenomenological model and results of numerical simulations for SSSR model. Panel (a): $A_{L,3}$ and $A_{R,3}$.  Panel (b): $\langle \beta_{L} \rangle_V$ and $\langle \beta_{R} \rangle_V$. Panel (c): $\langle |\mathbf{P}_R| \rangle_V$.
     }
     \label{fig:semi1}
\end{figure*}

\begin{figure*}
    \centering
    \subfigure[]{\includegraphics[width=0.49\textwidth]{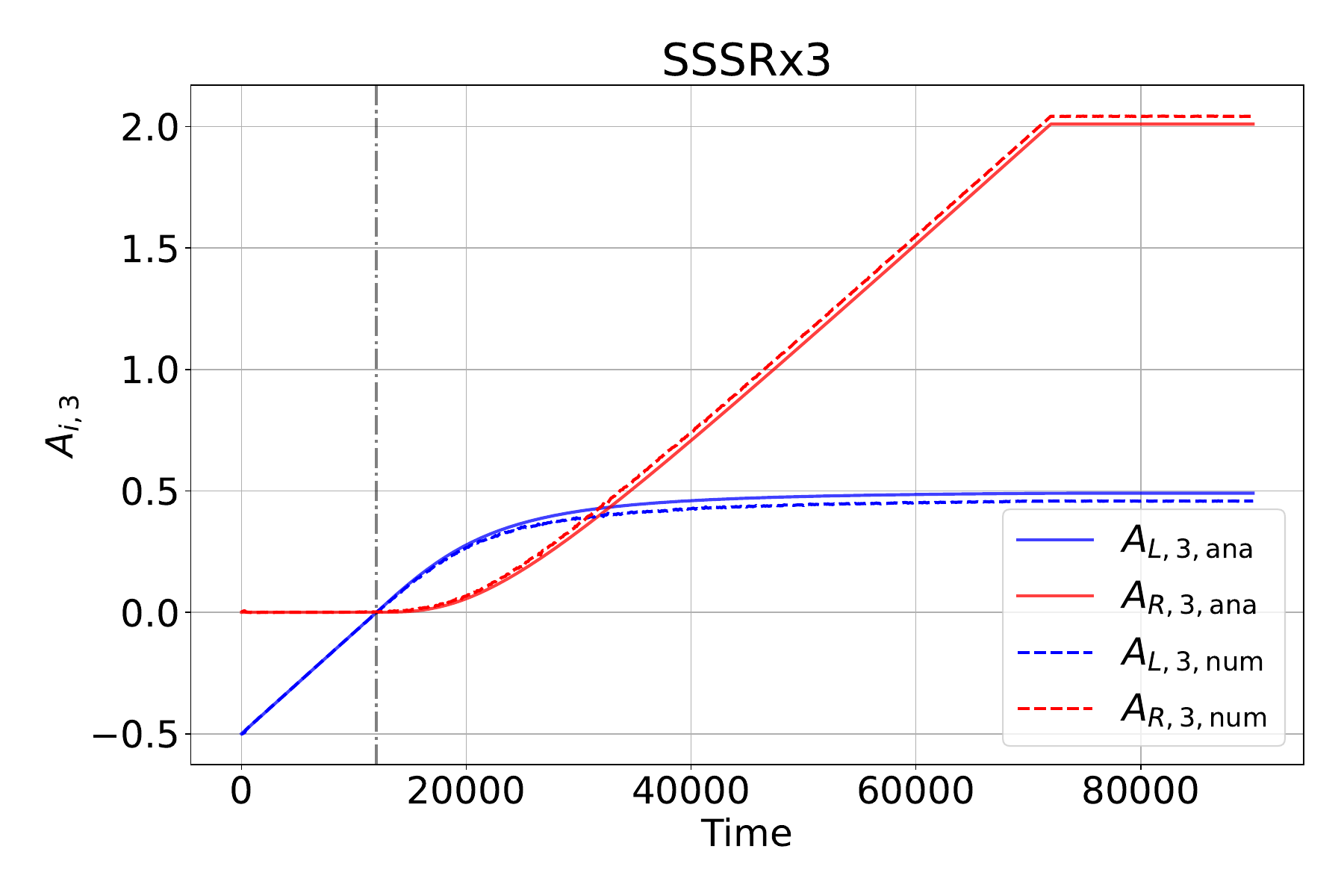}
         \label{subfig:semi_Ax3}
     }
     \subfigure[]{\includegraphics[width=0.49\textwidth]{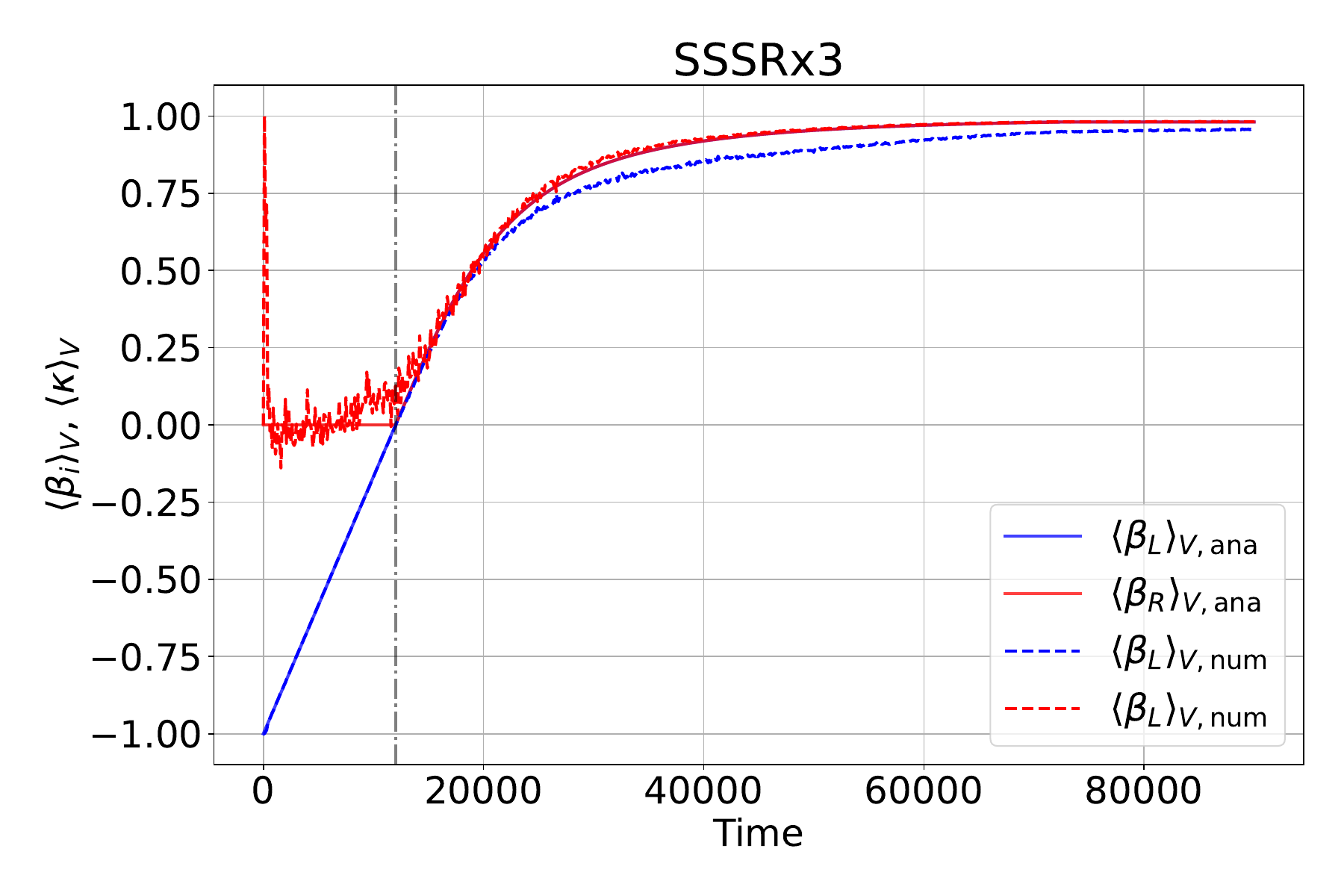}
         \label{subfig:semi_betax3}
     }
     \subfigure[]{\includegraphics[width=0.49\textwidth]{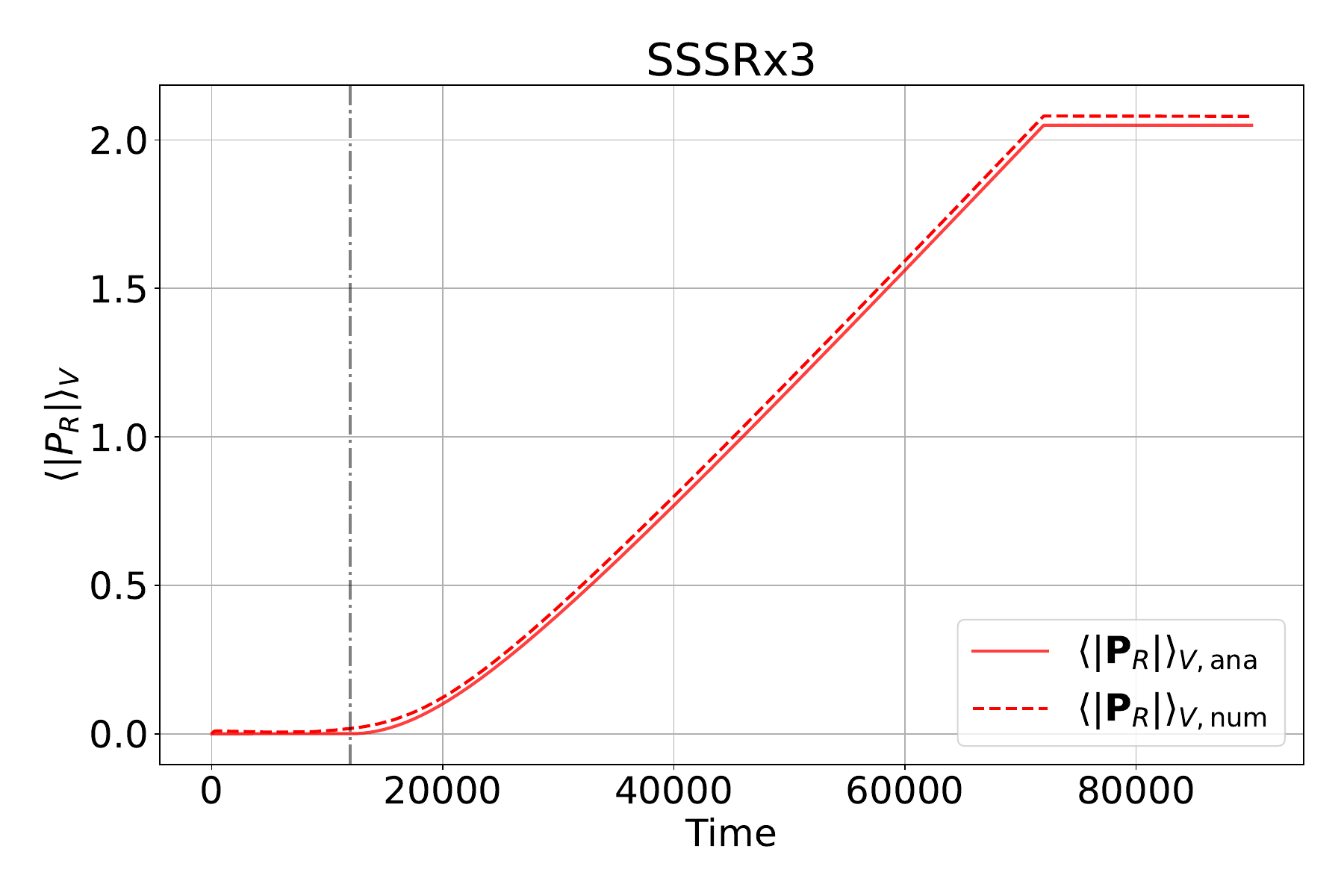}
         \label{subfig:semi_PRx3}
     }
     \caption{
     Same as Fig.~\ref{fig:semi1} but for SSSRx3 model.
     }
     \label{fig:semix3}
\end{figure*}

Based on the results of our quasi-homogeneous analysis presented above, we develop a phenomenological model for the quasi-steady evolution of FFC found in SSSR model. We assume that right-going neutrinos are fully depolarized in the phase at $t<t_m$, i.e., $\langle |\mathbf{P}_R| \rangle_V=0$ and $\langle \beta_R \rangle_V=0$ (although one can analytically approximate its value from the linear phase\footnote{\liu{In the linear phase, the z-component and norm grow linearly as $P_{R,3}(t)=|\mathbf{P}_R(t)|=C_Rt$, while the flavor coherence grows exponentially, $|\mathbf{P}^T_R|=\sqrt{P_{R,1}(t)^2+P_{R,2}(t)^2}\sim\mathrm{e}^{\mathrm{Im}[\omega(P_{R,3})]t}$. The saturation of the linear phase is marked by the point at which the magnitude of flavor coherence has grown to be a great fraction of the norm of the polarization vector. Henceforth the saturation is reached at smaller $|\mathbf{P}_R|$ (equivalently, smaller injection amounts) with lower $C_R$. In the non-linear phase for $t<t_m$, $\mathbf{P}_R$ is depolarized in the xy plane, which makes $|\mathbf{P}_R|$ cease evolving.}}). On the other hand, for left-going neutrinos, $\langle |\mathbf{P}_L| \rangle_V$ is constant in time. The time evolution of $\langle \beta_L \rangle_V$ can be obtained from the QKE for the zeroth angular moment of $\mathbf{A}_0(\equiv\mathbf{A}_L+\mathbf{A}_R)$,
\begin{equation}
\partial_t \mathbf{A}_0=(0,0,C_R(t)+C_L(t)),
\label{eq:zerothAevo_1}
\end{equation}
which can be solved analytically. \liu{In the analysis for SSSR model, $C_L$ is set to zero.} The solution can be expressed as
\begin{equation}
    \begin{split}
    &A_{0,1}(t)=A_{0,1}(0), \\
    &A_{0,2}(t)=A_{0,2}(0), \\
    &A_{0,3}(t)=A_{0,3}(t_m)+(t-t_m)C_R=(t-t_m)C_R.
    \end{split}
    \label{eq:anaA0solu}
\end{equation}
By using Eq.~\ref{eq:anaA0solu} and a factorized relation (Eq.~\ref{eq:factorize_beta}), the time evolution of $\langle \beta_L \rangle_V$ can be expressed analytically as,
\begin{equation}
    \langle\beta_L\rangle_V=\frac{(t-t_m)C_R}{\langle|\mathbf{P}_L|\rangle_V}.
    \label{eq:betaL_phase1}
\end{equation}

At the time $t \ge t_m$, on the other hand, we adopt a condition of
\begin{equation}
    \langle\beta\rangle_V\equiv\langle\beta_L\rangle_V=\langle\beta_R\rangle_V.
    \label{eq:betaLeqR}
\end{equation}
Assuming the factorization Eq.~\ref{eq:factorize_beta}, $A_{0,3}$ can be expressed as,
\begin{equation}
    A_{0,3}=\langle|\mathbf{P}_L|\rangle_V\langle\beta\rangle_V
    +\langle|\mathbf{P}_R|\rangle_V\langle\beta\rangle_V.
    \label{eq:A03_fact}
\end{equation}
By using Eqs.~\ref{eq:betaLeqR}~and~\ref{eq:A03_fact}, $\langle\beta\rangle_V$ can be expressed as
\begin{equation}
    \langle\beta\rangle_V=\frac{(t-t_m)C_R}{\langle|\mathbf{P}_L|\rangle_V+\langle|\mathbf{P}_R|\rangle_V}.
    \label{eq:semi1}
\end{equation}
It should be noted that $\langle|\mathbf{P}_R|\rangle_V$ evolves with time at $t>t_m$ for SSSR model, indicating that Eq.~\ref{eq:semi1} is still incomplete as an analytic expression for the dynamical evolution. By using Eqs.~\ref{eq:dt_PRnorm_1}~and~\ref{eq:semi1}, we can express the time evolution equation for $\langle|\mathbf{P}_R|\rangle_V$ as
\begin{equation}
    \partial_t\langle|\mathbf{P}_R|\rangle_V=\frac{(t-t_m)C_R}{\langle|\mathbf{P}_L|\rangle_V+\langle|\mathbf{P}_R|\rangle_V}C_R,
    \label{eq:dPRdt_closePR}
\end{equation}
which can be solved analytically as,
\begin{equation}
    \langle |\mathbf{P}_R| \rangle_V = -\langle|\mathbf{P}_L|\rangle_V + \sqrt{\langle |\mathbf{P}_L| \rangle_V^2 + \left( t^2 - 2 t_m t \right) C_R^2 + 2 F_{int}},
    \label{eq:PRnorm_anasol}
\end{equation}
where $F_{int}$ represents an integral constant. By using the condition of $\langle |\mathbf{P}_R| \rangle_V=0$ at $t=t_m$, we can obtain $F_{int}$ as,
\begin{equation}
     F_{int} =  \frac{t_m^2}{2} C_R^2.
     \label{eq:coef_PRdifsol2}
\end{equation}
Eqs.~\ref{eq:betaLeqR},~\ref{eq:semi1}, and \ref{eq:PRnorm_anasol} (with Eq.~\ref{eq:coef_PRdifsol2}) corresponds to the analytic solution for the time evolution of all three dynamical variables ($\langle\beta_{L,R}\rangle_V$ and $\langle |\mathbf{P}_R| \rangle_V$) in our phenomenological model. In Fig.~\ref{fig:semi1}, we compare the analytic solution with the simulation result for SSSR model. As can be clearly seen in the figure, our phenomenological model shows good agreement with the numerical result, indicating that the essential trend is well captured by the phenomenological model.

It would be worthwhile to discuss errors and limitations in the phenomenological model. First, $\langle|\mathbf{P}_R|\rangle_V$ for $t<t_m$ is assumed to be zero, but it should be finite in real cases. It could be estimated by a detailed analysis of the linear phase, but the magnitude would depend on $C_R$, indicating that there are no universal values. Second, our phenomenological model is developed based on an assumption that the timescale of FFC is much shorter than $1/C_R$. However, when $\langle\beta_{L/R}\rangle_V$ approaches unity, FFC tends to be less vigorous, implying that the quasi-steady condition would be no longer valid. We can assess the error by adopting our phenomenological model to SSSRx3 model, in which $\nu_e$ injection lasts three times as long as that in SSSR one. The result is displayed in Fig.~\ref{fig:semix3}. As shown in the figure, $\langle\beta_{L}\rangle_V$ has a relatively large difference between the phenomenological model and numerical simulation. It should be mentioned, however, that the error is within a few percents, indicating that the phenomenological model has sufficient ability to capture the qualitative trend of its quasi-steady evolution.

Finally, we show that our phenomenological model can provide an approximate solution for $\langle \mathbf{P}_L\times\mathbf{P}_R \rangle_V$, which is the driving term of flavor mixing for the spatial averaged neutrino distributions (see Eq.~\ref{eq:QKE_A}). As can be seen in the Fourier analysis of QKE, it corresponds to a convolution term for the spatial wave number ($K$). This suggests that the term determines the feedback to homogeneous mode by non-linear couplings among finite $K$ modes. From Eqs.~\ref{eq:QKE_A}~and~\ref{eq:factorize_beta}, the time derivative of $\langle\beta_{L/R}\rangle_V$ can be written as
\begin{equation}
    \begin{split}
        \partial_t\langle\beta_R\rangle_V&=\partial_t\frac{A_{R,3}}{\langle|\mathbf{P}_R|\rangle_V}  \\      
        &=\frac{2\langle \mathbf{P_L}\times\mathbf{P_R}\rangle_{V,3}+C_R(t)}{\langle|\mathbf{P}_R|\rangle_V}-\frac{\langle\beta_R\rangle_V^2C_R(t)}{\langle|\mathbf{P}_R|\rangle_V},\\
        \partial_t\langle\beta_L\rangle_V&=\partial_t\frac{A_{L,3}}{\langle|\mathbf{P}_L|\rangle_V}  \\      
        &=\frac{-2\langle \mathbf{P_L}\times\mathbf{P_R}\rangle_{V,3}+C_L(t)}{\langle|\mathbf{P}_L|\rangle_V}-\frac{\langle\beta_L\rangle_V^2C_L(t)}{\langle|\mathbf{P}_L|\rangle_V}.
    \end{split}
    \label{eq:EOM_beta}
\end{equation}
By using a condition of $\partial_t\langle\beta_R\rangle_V=\langle\beta_R\rangle_V=0$ at $t<t_m$ in our phenomenological model, we can obtain $\langle \mathbf{P}_L\times\mathbf{P}_R \rangle_V$ from Eq.~\ref{eq:EOM_beta} as
\begin{equation}
    \langle \mathbf{P_L}\times\mathbf{P_R}\rangle_{V,3}\approx-\frac{C_R}{2}.
    \label{eq:PLxPR_1}
\end{equation}
For the phase at $t>t_m$, on the other hand, it can be expressed by using Eqs.~\ref{eq:betaLeqR}~and~\ref{eq:EOM_beta} as
\begin{equation}
    \langle \mathbf{P_L}\times\mathbf{P_R}\rangle_{V,3}\approx-\frac{1}{2}\frac{(1-\langle\beta\rangle_V^2)C_R}{1+\langle|\mathbf{P}_R|\rangle_V/\langle|\mathbf{P}_L|\rangle_V}.
    \label{eq:PLxPR_2}
\end{equation}
We confirm that Eqs.~\ref{eq:PLxPR_1}~\ref{eq:PLxPR_2} are in good agreement with numerical simulations (although numerical or statistical noises are present). One thing we learn from this analysis is that $\nu_e$ injection induces a correlation between $\liu{\beta}_L$ and $\liu{\beta}_R$, which dictates a quasi-steady evolution of the system. For a comprehensive understanding of the detail of mode coupling, we need more detailed studies, which will be discussed in forthcoming papers.

\section{Summary and discussion}
\label{sec:sumdis}
FFC can occur over very short spatial and temporal scales compared to those involved in astrophysical systems. It is currently impossible to carry out global neutrino transport simulations in astrophysical environments while adequately resolving such short scales. One of the most common prescriptions to include effects of micro-scale processes into large-scale simulations is through subgrid modeling. In this approach, coarse-grained variables are used to represent global properties of the system, while the feedback from small-scale structure is approximated one way or another. Local simulations are powerful tools to investigate how small-scale flavor conversions build up large-scale structures in neutrino radiation fields, which offers a clue to develop subgrid models. In a series of our papers \cite{PhysRevLett.129.261101,PhysRevD.107.063033,PhysRevD.107.103022}, we proposed that the disappearance of ELN-XLN crossings in neutrino angular distributions is a reasonable condition to determine asymptotic states of FFCs. This prescription has been corroborated by independent research groups through their own local QKE simulations \cite{xiong2023evaluating,richers2024asymptoticstatepredictionfastflavor,george2024evolutionquasistationarystatecollective}.

Very recently, however, the authors in \cite{fiorillo2024fastflavorconversionsedge} demonstrated that FFC can alter spatially-averaged neutrino radiation fields even without ELN-XLN crossings in spatially-averaged neutrino angular distributions, if neutrinos are injected with time. This posed a challenge for developments of subgrid models for FFCs. In fact, the canonical phenomenological model would lead to qualitatively different asymptotic states from those obtained from numerical simulations. In this paper, we delve into the anomaly of FFC reported in \cite{fiorillo2024fastflavorconversionsedge} by carrying out numerical simulations systematically with detailed analytic arguments based on quasi-homogeneous analysis (Sec.~\ref{subsec:quahomoana}). This analysis provides key ingredients to develop a phenomenological model, in which we can determine the time evolution of the system analytically. Our major findings and conclusions are summarized as follows.

\begin{enumerate}
\item Our numerical simulations reproduced SR model in \cite{fiorillo2024fastflavorconversionsedge}. On the other hand, we found that the relation of $A_{L\liu{,}3}  \sim A_{R\liu{,}3} $ appeared in the late phase is not an intrinsic property of quasi-steady evolution of FFC. In fact, the time evolution of $A_{L,3}$ and $A_{R,3}$ is remarkably different from each other for cases with slower neutrino injections (SSSR model).
\item In SSSR model, on the other hand, we found that $\langle \beta_{L} \rangle_V \sim \langle \beta_{R} \rangle_V$ (see Eq.~\ref{eq:betadef} for definition of $\beta$). The trend can be understood by the time evolution property of $\langle \kappa \rangle_V$ (see Eq.~\ref{eq:tevokappa}) with quasi homogeneous model (Sec.~\ref{subsec:quahomoana}). This interpretation is also supported by comparing to SSSRs1 model, in which $\langle \kappa \rangle_V$ is constant after $\nu_e$ injection is swapped from the right- to the left-going direction (see Fig.~\ref{fig:SSSRs1_beta}).
\item In Sec.~\ref{sec:semi_ana}, we demonstrate that the essential feature of quasi-steady evolution of FFCs can be determined analytically (see Fig.~\ref{fig:semi1}). One of the key conditions in the phase at $t<t_m$ is that the right-going neutrinos are fully depolarized, which is consistent with erasing ELN-XLN angular crossings. At $t>t_m$, we assume that $\langle \beta_{L} \rangle_V$ is equal to $\langle \beta_{R} \rangle_V$, and that $\langle |\mathbf{P}_{L/R}| \beta_{L/R} \rangle_V$ can be factorized (see Eq.~\ref{eq:factorize_beta}). These conditions are supported by our quasi-homogeneous analysis with numerical simulations.
\item Our phenomenological model also allows us to express the relation between $\langle \mathbf{P_L}\times\mathbf{P_R}\rangle_{V,3}$ and $C_R$ in SSSR model (see Eqs.~\ref{eq:PLxPR_1}~and~\ref{eq:PLxPR_2}). We note that $\langle \mathbf{P_L}\times\mathbf{P_R}\rangle_{V,3}$ corresponds to a convolution term of spatial wave number, which represents non-linear feedback from small-scale structures of flavor conversions. This provides a clue to study the correlation between \liu{the two beams} $\mathbf{P_L}$ and $\mathbf{P_R}$, although the detailed investigation is postponed to future work.
\end{enumerate}

Before closing the paper, an important remark is worth making. As we showed in the present work, the information on spatially-averaged (or homogeneous part of) polarization vectors of neutrinos is insufficient to determine the overall trend of neutrino radiation field. We find, on the other hand, quasi homogeneous quantities such as $\langle |\mathbf{P}| \rangle_V$, $\langle \beta \rangle_V$, and $\langle \kappa \rangle_V$ offer complementary information. It is interesting to consider how we can include them into subgrid modeling for global neutrino transport simulations, whose detailed study is deferred to future work.

\begin{acknowledgments}
H.N. is supported by Grant-in-Aid for Scientific Research (23K03468) and also by the NINS International Research Exchange Support Program.
M.Z. is supported by Grant-in-Aid for JSPS Fellows (Grant No. 22KJ2906) and JSPS KAKENHI Grant Number JP24H02245.
L.J. is supported by a Feynman Fellowship through LANL LDRD project number 20230788PRD1.
R.A. is supported by JSPS Grants-in-Aid for Scientific Research (JP24K00632).
S.Y. is supported by Grant-in-Aid for Scientific Research (21H01083).
S.Y. is supported by the Institute for Advanced Theoretical and Experimental Physics,
Waseda University, and the Waseda University Grant for Special Research Projects (project No. 2023C-141, 2024C-56, 2024Q-014).
Numerical computation in this work was carried out at the Yukawa Institute Computer Facility.
This work is also supported by the HPCI System Research Project (Project ID:240079).
\liu{We thank Damiano F.G. Fiorillo for valuable comments.}
\end{acknowledgments}
\bibliography{refe}
\bibliographystyle{apsrev4-2}

\end{document}